\documentclass[11pt]{article}

\usepackage[utf8]{inputenc}
\usepackage{amsmath}
\usepackage{amssymb}
\usepackage{mathtools}
\usepackage{hyperref}
\hypersetup{colorlinks=true}
\usepackage{graphicx}
\usepackage{grffile}
\usepackage{cite}

\usepackage{algpseudocode}
\usepackage{algorithm}

\usepackage{caption} 

\usepackage{multirow}
\usepackage[table,xcdraw]{xcolor}

\newcommand{\figwidth}{0.62\textwidth} 

\newcommand{\R}{\mathbb{R}}

\newcommand{\U}{\mathcal{U}}
\DeclareMathOperator*{\argmin}{arg\,min}

\newcommand{\norm}[2]{{\|{#2}\|}_{#1}}
\def\code#1{\texttt{#1}}
\newcommand{\s}{\text{\,s}}
\newcommand{\Hz}{\text{\,Hz}}

\newcommand{\ifourier}[1]{\mathcal{F}^{-1}\{#1\}}

\title{SRMD: Sparse Random Mode Decomposition\\}
\author{Nicholas Richardson, Hayden Schaeffer, and Giang Tran}
\date{ }

\begin{document}

\maketitle

\begin{abstract}
Signal decomposition and multiscale signal analysis provide many useful tools for time-frequency analysis. We proposed a random feature method for analyzing time-series data by constructing a sparse approximation to the spectrogram. The randomization is both in the time window locations and the frequency sampling, which lowers the overall sampling and computational cost. The sparsification of the spectrogram leads to a sharp separation between time-frequency clusters which makes it easier to identify intrinsic modes, and thus leads to a new data-driven mode decomposition. The applications include signal representation,  outlier removal, and mode decomposition. On benchmark tests, we show that our approach outperforms other state-of-the-art decomposition methods.  
\end{abstract}

\section{Introduction}
Time-frequency analysis is an important tool for analyzing and leveraging information from signals. Various time-frequency approaches lead to a particular decomposition of signals into important time-varying features which can illuminate intrinsic behaviors, be used for analytics, or assist in smoothing the signal. For example, the classical short-time Fourier transform (STFT) extracts a time-varying representation by localizing the signal in time using a finite window length and applying the Fourier transform to each localized segment. Since the window length is fixed, the resulting analysis leads to a uniform time-frequency resolution. The continuous wavelet transform (CWT) constructs a representation of the signal using a variable time window by scaling the mother wavelet and thus leads to a multiscale representation of the signal. Modern techniques focus on addressing several issues with signal decompositions, specifically, data-driven representations that better reflect the signal's intrinsic behavior, sharpen/localize the spectrogram, and are robust to noise, outliers, or non-uniform sampling.

 The empirical mode decomposition (EMD) \cite{huang1998empirical} is an adaptive time-frequency method for analyzing and decomposing signals and has been shown to be useful in a wide range of signal processing applications. In particular, EMD decomposes a given signal into intrinsic mode functions (IMFs) which carry information at varying frequency scales by detecting local extrema and estimating upper and lower envelopes. This effectively partitions the spectrum into certain frequency bands, which are represented by the learned IMFs. EMD suffers from some problems including \textit{mode mixing}, i.e.\ the appearance of similar frequency information shared between distinct IMFs, and sensitivity to noise and sampling. The ensemble EMD (EEMD) \cite{wu2009ensemble} learns the IMFs using an ensemble of the given signal perturbed by random (Gaussian) noise. This helps to mitigate the mode mixing issue by leveraging results on EMD applied to white noise \cite{flandrin2004empirical}; however, the approximated signals often retain aspects of the noise and the perturbations may lead to a different IMF decomposition. In \cite{torres2011complete}, the complete EEMD with adaptive noise (CEEMDAN) added different (synthetic) noise to the stages of the decomposition which led to more stable results. Since EEMD-based approaches average over several applications of EMD, they often come with an increased cost. In \cite{hou2011adaptive},  the signal is represented using an IMF-like form and its time-varying parameters are optimized using a total variational penalty on the third derivative of the coefficients. The variational mode decomposition (VMD) \cite{dragomiretskiy2013variational} decomposes the signal into a sum of IMFs using an optimization problem (implemented by the split Bregman or ADMM method \cite{goldstein2009split}). The IMFs are obtained simultaneously within the optimization process and the resulting decompositions are more stable to noise than the standard EMD approaches.

The synchrosqueeze transform (SST) \cite{daubechies2011synchrosqueezed} improves over the CWT by calculating instantaneous frequencies and  ``squeezing'' them through a reassignment algorithm, namely, shifting them to the center of the time-frequency region \cite{auger2013time}. This leads to sharper time-frequency representations than the STFT and CWT, which are often limited by the finite sampling lengths and can create spectral smearing. In addition, the sharpening essentially prunes the unnecessary wavelet coefficients, thus leading to a sparser representation. Various SST-based methods have been proposed using other signal transforms for example the S-transform \cite{huang2015synchrosqueezing} and the wavelet packet transform \cite{yang2015synchrosqueezed}.  The empirical wavelet transform (EWT) \cite{gilles2013empirical} combines aspects of EMD with the wavelet transform. The main idea behind EWT is to partition the Fourier domain and build empirical wavelet filters from the segmented spectrum. This is done by identifying the local maxima of the amplitude in the Fourier domain and partitioning the regions to separate the maxima \cite{liu2019recent}. In \cite{gilles2014parameterless}, the authors propose a fast scale-space algorithm to automatically detect meaningful modes from a histogram without assuming the knowledge about the number of modes. The EWT was extended to two dimensions for applications in imaging and can be related to other wavelet-like transforms \cite{gilles20142d}.

In this work, we propose a signal representation and decomposition algorithm based on a sparse random feature approximation \cite{hashemi2021generalization} to the continuous short-time Fourier transform. The previous paragraph summarized several successful signal decomposition techniques, some of which are developed directly from a continuous transform (e.g. STFT, SST, CWT, etc.) or more generally a signal-frequency decomposition form (e.g. EMD and the related methods). In this work, we represent the signal using the continuous STFT and develop a randomized approach for approximation of the continuous integral formulation without needing to resolve the time-frequency domain. The methods are related to the random feature models, which are used in regression, classification, and more generally supervised learning. 

 Random feature models (RFMs) are a randomized nonparametric approximation used in interpolation and regression problems \cite{rahimi2007random, rahimi2008uniform, rahimi2008weighted}. It can be viewed as a kernel-based approach or as a class of artificial neural networks. Specifically, the standard RFM architecture consists of a two-layer fully connected neural network whose single hidden layer is randomized and not trained \cite{block1962perceptron, rahimi2007random, rahimi2008uniform, maass2004computational, moosmann2006randomized}. The only layer that is trained is the output layer thus yielding a linear training model. There is a wide range of theoretical results for RFMs used in interpolation or regression \cite{rahimi2008weighted, rudi2017generalization,li2019towards,weinan2020towards, hashemi2021generalization,mei2021generalization, bach2017equivalence, sriperumbudur2015optimal}. In \cite{rahimi2008uniform} using $N$ random features is shown to yield a uniform error bound of $\mathcal{O}(N^{-\frac{1}{2}}+m^{-\frac{1}{2}})$ for  target functions in a certain class when the RFM is trained using Lipschitz loss functions. For the $L^2$ loss, if the number of features scale like $N\sim \sqrt{m} \, \log{m}$ where $m$ is the number of data points, then the test error is bounded by $\mathcal{O}(m^{-\frac{1}{2}})$ \cite{rudi2017generalization}, see also \cite{li2019towards}. This result requires that the target function $f$ is in the associated reproducing kernel Hilbert space (RKHS) and some additional assumptions on the kernel. By analyzing the structure of the RFM with respect to the dimension $d$ and the parameters $N$ and $m$, results found in \cite{mei2021generalization, chen2021conditioning} showed that regression using the RFM often achieve their minimal risk in the overparameterized region, where the number of random features exceeds the number of data samples. 

One family of approaches to learn RFMs in the overparameterized regime is based on imposing a sparsity prior in the number of features. In \cite{hashemi2021generalization}, the $\ell^1$ basis pursuit denoising problem was used to obtain a low complexity RFM (measured in terms of the number of active random features), see also \cite{yen2014sparse}. The test error scales like the standard RFMs and improves when the target function has fast decay relative to the random features \cite{hashemi2021generalization, chen2021conditioning}. In \cite{saha2022harfe}, a hard ridge-based thresholding algorithm, called HARFE, was proposed to iteratively obtain sparser RFMs by solving a sparse ridge regression problem \cite{luedtke2014branch, mazumder2017subset, bertsimas2020sparse, hazimeh2020fast, xie2020scalable}. A random feature pruning method called the SHRIMP algorithm was proposed in \cite{xie2021shrimp}. The approach iteratively prunes an overparameterized RFM by alternating between a thresholding step and a regression step and was shown to perform well on both real and synthetic data. As noted in \cite{xie2021shrimp}, RFMs with sparsity priors can also be motivated by the lotto ticket hypothesis, which asserts the existence of small subnetworks within overparameterized neural network who match or improve on the accuracy of the full network \cite{frankle2018lottery}. 


\subsection{Our Contributions}
Some of our main algorithmic and modeling contributions are as follows:
\begin{itemize}
\item We propose a randomized sparse time-frequency representation, which extends and further develops the sparse random feature method \cite{hashemi2021generalization, saha2022harfe}. Specifically, our model allows for the use of compressive sensing-like techniques to spectrogram analysis with a less restrictive ``basis'', i.e.\ the random feature space. In addition, the randomization allows for non-equally spaced points in both time and frequency.
\item We show that our approach produces a mode decomposition with less mode mixing, better separation of modes, and fewer Gibbs phenomena than other state-of-the-art approaches. The sparsity prior sharpens the spectrogram which leads to a clearer separation between modes and allows for simple clustering of the time-frequency regions. 
\item Our method can also be used for outlier or corruption removal, in particular, removing nonlinear and highly correlated noise from the data (see Section~\ref{sec:astronomy}). One application is for data-assisted modeling for scientific discovery, where our approach can be used to guide one to a particular structure or waveform with prior templates or information.
\item Our approach does not require that the time series is obtained from equally spaced time points which makes it applicable for a wider range of datasets where standard approaches fail. In fact, the method does not depend strongly on the sampling process, unlike other state-of-the-art approaches. 
\end{itemize}

\section{Sparse Random Feature Representation for Time-Series Data}
The proposed method builds from the continuous STFT, that is, we represent a signal $f \in L^1([0,T])$ by
\begin{align*}
f(t) = \int_{-\infty}^\infty f(t)W(t-\tau) \text{d}\tau=  \int_{-\infty}^\infty \int_{-\infty}^\infty\, \alpha(\omega, \tau) \, W(t-\tau)\, \exp{(i\omega t)}\, \text{d}\omega \text{d}\tau,
\end{align*}
where $\alpha$ is the transform function and $W$ is a (positive) window function such that
$\int_{-\infty}^\infty W(t-\tau) d\tau=1$.
The assumptions are that the transform function is band-limited, i.e.\ $\alpha(\omega, \tau)=0$ for all $|\omega|>B$ and that for a fixed $\tau$ the support of $\alpha(\omega, \tau)$ is small. Note that since $f \in L^1([0,T])$, the transform is bounded, i.e.\ there exists an $M>0$ such that $|\alpha(\omega, \tau)|\leq M$ for all $(\omega, \tau)$. Let $N$ be the number of total random features used in the approximation of the integral above. In particular, using the RFM, we approximate the integrals using $N=N_1N_2$ random features where $N_1$ is the number of random frequencies and $N_2$ is the number of random windows:

\begin{align*}
f(t) = \int_{-\infty}^\infty \int_{-B}^B\, \alpha(\omega, \tau) \, W(t-\tau)\, \exp{(i\omega t)}\, \text{d}\omega \text{d}\tau\approx  \sum_{k_2=1}^{N_2}  \sum_{k_1=1}^{N_1} c_{k_1,k_2} \, W(t-\tau_{k_2})\, \exp{(i\, \omega_{k_1} \, t)}
\end{align*}
where $\left\{ \omega_{k_1} \right\}_{k_1\in[N_1]}$ and $\left\{ \tau_{k_2} \right\}_{k_2\in[N_2]}$ are independent of each other and are drawn i.i.d. $\omega_{k_1} \sim \U[0,B]$ and $\tau_{k_2} \sim \U[0,T]$. The goal is to learn a representation of the target signal $f$ using $m$ sampling points $\{ t_\ell\}_{\ell \in [m]}~\subset~[0,T]$. The sampling points can either be equally spaced in time or can be drawn i.i.d. from a probability measure $\mu(t)$ along the interval $[0,T]$. The given output measurements $y_{\ell}$ are $y_{\ell} = f(t_{\ell})+e_{\ell}$, where the noise or outliers $\{ e_{\ell}\}_{\ell \in [m]}$ are either bounded by constant $E>0$, i.e.\ $|e_{\ell} |\leq E$ for all $ \ell \in [m]$, or are random Gaussian. Note that if $e_{\ell}\sim \mathcal{N}({0},\sigma^2)$ and if $m\geq 2 \log\left(\delta^{-1} \right)$, then the noise terms $e_{\ell}$ are bounded by $E=2\sigma$ for all $\ell \in[m]$ with probability exceeding $1-\delta$.

We reindex $(k_1,k_2)$ so that the $N$ random feature functions take the form $$\phi_j(t):=W(t-\tau_j)\, \exp{(i\, \omega_j \, t)}$$ with the new index $j\in[N]$ and thus the approximation becomes 
\begin{equation}
f(t)\approx \sum_{j=1}^N c_j \, \phi_j(t),
\end{equation}
where the coefficients $c_j$ have also been reindexed. The training problem becomes learning coefficients $c_j$ so the approximation $\sum_{j=1}^N c_j \, \phi_j(t_{\ell})$ is close to the given data $y_\ell$.  Let $\mathbf{A} \in \mathbb{C}^{m \times N}$ be the random feature matrix whose elements are defined as
$a_{\ell,j} = \phi_j(t_\ell)$,  $\mathbf{c}=[c_1, \ldots, c_N]^T$ and $\boldsymbol{y}=[y_1, \ldots, y_m]^T$. By assumption, the time-frequency representation is sparse, so we learn $\mathbf{c}$ by solving an $\ell^1$ regularized least squares problem.  Following \cite{hashemi2021generalization}, a sparse random feature model can be trained with the $\ell^1$ basis pursuit denoising problem \cite{candes2006near, cai2009recovery, foucart2013sparse}:
\begin{equation}\label{eq:BP}
\mathbf{c}^\sharp = \argmin_{\mathbf{c}\in\mathbb{C}^N}\   \|\mathbf{c}\|_1 \quad \text{s.t. }\quad \|\mathbf{A}\mathbf{c}- \boldsymbol{y}\|_2\leq \eta\sqrt{m},
\end{equation}
where $\eta$ is a user-defined parameter that is related to the noise bound $E$. It can be shown that certain random feature matrices are well-conditioned to sparse regression when trained using Equation \eqref{eq:BP} \cite{hashemi2021generalization,chen2021conditioning}. When the input data is contaminated by large noise (such as the gravitational distortion data in Section \ref{sec:astronomy}), we solve the unconstrained $\ell^1$ optimization problem (LASSO) \cite{tibshirani1996regression,hastie2019statistical}:  \begin{equation}\label{eq:LASSO}
\mathbf{c}^\sharp = \argmin_{\mathbf{c}\in\mathbb{C}^N}\  \lambda  \|\mathbf{c}\|_1+  \frac{1}{2m} \|\mathbf{A}\mathbf{c}- \boldsymbol{y}\|_2^2.
\end{equation}
Although the unconstrained $\ell^1$ optimization problem is equivalent to the basis pursuit denoising problem under a mapping between $\lambda>0$ and $\nu>0$ \cite{foucart2013sparse}; empirical tests for this particular application showed that the LASSO form was more forgiving while tuning parameters. We expect that this is a consequence of the specific choice of algorithms more than the specific formulation.

\subsection{Sparse Random Feature Representation Algorithm}
In the algorithm, we replace the complex exponential by a sine function with a random phase $\psi_j$ and the window function is defined by the Gaussian with a fixed variance $\Delta^2$:
\begin{equation}
\phi_{j}(t) =\exp\left(-\frac{(t-\tau_j)^2}{2\Delta^2}\right)\sin(2\pi  \omega_j\, t +\psi_j), 
\end{equation}
where $\tau_j \sim \U(0,T), \omega_j \sim \U(0,\omega_{\max}),$ and $\psi_j \sim \U(0,2\pi).$ Given a set of time points $\left\{ t_\ell \right\}_{\ell \in[m]}$, we define the random short-time sinusoidal feature matrix $\mathbf{A}=[a_{\ell,j}]\in \mathbb{R}^{m\times N}$ by
$$a_{\ell,j} = \phi_j(t_\ell)= \exp\left(-\frac{(t_\ell-\tau_j)^2}{2 \Delta^2}\right)\, \sin(2\pi \omega_j t_\ell +\psi_j).$$
While the standard approaches assume that data is obtained from an evenly spaced time series, we do not place any restrictions on the sampling of the time points (except that they are distinct). This is an important distinction compared to other signal decomposition approaches. In particular, we optimize the coefficients $\mathbf{c}$ using a sparse optimization problem, Equation \eqref{eq:BP}, with the random short-time sinusoidal feature matrix, which can be shown to lead to well-conditioned training even in the low data limit \cite{hashemi2021generalization, chen2021conditioning}. As an added benefit, the random sampling of time points reduces the computational and storage cost, which depends on the number of samples and the number of features. The reconstruction algorithm is summarized in Algorithm \ref{alg:reconstruction}.

\bigskip

\begin{algorithm}[htp]
\caption{Sparse Random Feature Representation for Time-Series Data}\label{alg:reconstruction}
\begin{algorithmic}[t]
\Require Samples $\{(t_{\ell},y_{\ell})\}_{\ell =1}^m$, number of random features $N$, maximum frequency $\omega_{\max}$, window size $\Delta$, noise level $r\in \,[0,1]$. Let $\mathbf{y} = [y_1,\ldots, y_m]^T$.
\Ensure

\State {\bf Draw} random time shifts, frequencies, and phases (independent of the data) \[\{\tau_j,\omega_j,\psi_j\}_{j=1}^N \sim \U[0,T]\times  \U[0,\omega_{\max}]\times \U[0,2\pi]\]
\State {\bf Construct} the random short-time sinusoidal feature matrix
\[ \mathbf{A}=[\phi_j(t_\ell)] = \left [ \exp\left(-\frac{(t_\ell-\tau_j)^2}{2 \Delta^2}\right)\, \sin(2\pi \omega_j t_\ell +\psi_j)\right] \in \R^{m\times N}.\]

\State {\bf Solve:}\quad $\mathbf{c}^\sharp = \argmin\limits_{\mathbf{c}\in\R^{N}} \norm{1}{\mathbf{c}} \text{ s.t. } \norm{2}{\mathbf{A}\mathbf{c}-\mathbf{y}} < \sigma=r\norm{2}{\mathbf{y}}$. 

\State {\bf Output:} Coefficient vector $\mathbf{c}^\sharp$ and the sparse random feature representation for the time-series:    
\[f^{\sharp}(t) =\sum\limits_{j=1}^N \ c_j^\sharp \ \phi_j(t).\]

\end{algorithmic}

\end{algorithm}

\subsection{Sparse Random Mode Decomposition (SRMD)}
In this section, we discuss how to utilize the learned coefficients $\mathbf{c}^\sharp$ to decompose the signal into meaningful modes. It is based on the observation that the sparse optimization extracts a sparse time-frequency representation, which has the added benefit of forming a simple decomposition due to the sharpening of the spectrogram. Specifically, we first collect all pairs $(\tau_j,\omega_j)$ corresponding to the non-zero learned coefficient $c^\sharp_j$, denoting the support set by
\[S :=\{(\tau_j,\omega_j)\mid j \in [N], \ c^\sharp_j\not =0\}.\]
We then partition $S$ into clusters using the clustering method DBSCAN. The learned coefficients are grouped based on those clusters and these groups define the corresponding IMFs. An advantage of DBSCAN is that we do not need to specify the number of extracted modes, which is useful for blind source separation. We discuss in Section \ref{sec:numerical} how to merge modes together if the number of modes is given. The sparse random mode decomposition (SRMD) algorithm is summarized in Algorithm~\ref{alg:decomposition}.

\begin{algorithm}[htp]
\caption{SRMD for Time-Series Data}\label{alg:decomposition}
\begin{algorithmic}[t]
\Require Samples $\{(t_{\ell},y_{\ell})\}_{\ell =1}^m$, number of random features $N$, maximum frequency $\omega_{\max}$, window size $\Delta$, noise level $r\in \,[0,1]$, $\code{frqscale}\in\R^+$, DBSCAN hyperparameters $\varepsilon$ and $\code{min\_samples}$. Let $\mathbf{y} = [y_1,\ldots, y_m]^T$.
\Ensure

\State {\bf Apply} Algorithm 1 to 
obtain $S=\{(\tau_j,\omega_j)\mid j\in [N], \ c^\sharp_j\not =0\}.$

\State {\bf Scale} input points to obtain $\widehat{S} =\{(\tau_{j},\widehat{\omega}_j)\mid j\in [N],\ c^\sharp_j\not =0, \ \widehat{\omega}_j = {\code{frqscale}} \cdot\omega_{j} \}.$

\State Partition $\widehat{S}$ into clusters $S_1,\ldots, S_K$ using DBSCAN. Let

\[I_k=\{j\in [N] \mid (\tau_j,\widehat{\omega}_j) \in S_k\},\quad k=1,\ldots K.\]
\State {\bf Output:} $K$ modes
\[y_k(t): = \sum\limits_{j\in I_k}\ c^\sharp_j \  \phi_j(t),\quad k=1,\ldots K.\]
\end{algorithmic}

\end{algorithm}

\section{Numerical Experiments}\label{sec:numerical}
In this section, we verify the applicability and consistency of SRMD on five decompositions and signal representation examples, including three challenging synthetic time-series from \cite{daubechies2011synchrosqueezed, dragomiretskiy2013variational} and two real time-series (musical and gravitational distortion datasets). In Section \ref{sec:discontinuity}, we display the learned coefficients obtained from Algorithm~\ref{alg:decomposition} in the time-frequency space (spectrogram) and plot the corresponding modal decomposition (i.e.\ clusters) using DBSCAN. While more sophisticated clustering algorithms can be used, one of the benefits of our algorithm is that it produces a sparse spectrogram that can be more easily clustered when the original signal has sharp time-frequency bands or groups. We also compare our approach with some of the state-of-the-art intrinsic mode decomposition methods, including the EMD \cite{huang1998empirical}, Ensemble EMD (EEMD) \cite{wu2009ensemble}, Complete EEMD with Adaptive Noise (CEEMDAN) \cite{torres2011complete}, Empirical Wavelet Transform (EWT) \cite{gilles2013empirical}, and Variational Mode Decomposition (VMD) \cite{dragomiretskiy2013variational}. Additionally, in Section \ref{sec:intersecting}, we compare our method to the spectrogram produced by the short-time Fourier transform (STFT), Continuous Wavelet Transform (CWT), and SST \cite{daubechies2011synchrosqueezed}. In Section \ref{sec:VMD_eq32}, we investigate the robustness of our method with respect to noise and the stability of our reconstruction and decomposition results with respect to the randomness of the random feature. In particular, we generate a musical time series using two modes (flute and guitar) and thus can compare our method to the ground truth modes when the data has unknown acquisition and background noise. As an application to data-assisted scientific discovery, we show that our approach can extract an approximate waveform for the merging event of two black holes without the need for templates (based on the LIGO dataset). More experiments on benchmark examples are shown in the Appendix.

All tests were performed using Python and our codes as well as the corresponding audio files for the musical example are available on GitHub\footnote{\url{https://github.com/GiangTTran/SparseRandomModeDecomposition}}. PyEMD \cite{pele2008, pele2009}, was used to test EMD and the related methods, while $\mathtt{ewtpy}$ and $\mathtt{vmdpy}$ \cite{carvalho2020evaluating} were used to test EWT and VMD, respectively. The hyperparameters used in all methods are chosen to optimize their resulting outputs and are based on the suggested parameters from the papers' methods or documentation. More precisely, for EMD, EEMD, and CEEMDAN, the threshold values on standard deviation, on energy ratio, and on scaled variance per IMF check are $\mathtt{std\_thr} = 0.1$, $\mathtt{energy\_ratio\_thr} = 0.1$, and $\mathtt{svar\_thr} = 0.01$, respectively. For EEMD and CEEMDAN, the number of noise-perturbed ensemble trials is set to $trials = 100$. For VMD, we set the balancing parameter of the data-fidelity constraint $alpha = 50$, the time-step of the dual ascent $tau = 1$, the convergence tolerance $tol = 10^{-6}$, and all frequencies $\omega$ are initialized randomly. 

For the hyperparameters used in SRMD, we need to choose $(\omega_{\max},\Delta,N)$ to generate the random feature matrix $\mathbf{A}$, the noise-level parameter $\eta$ for the SPGL1 algorithm, and $\code{frqscale}$, $\code{min\_samples}$, and $\varepsilon$ for the DBSCAN. We used the following values unless stated otherwise. The maximum possible frequency $\omega_{\max}$ is set to the Nyquist rate, $\omega_{\max} = \dfrac{m}{2T}$, the  
window size is set to $\Delta=0.1$, and the number of randomly generated features $N$ can range from $5m$ to $50m$ depending on the complexity of the signal. The basis pursuit parameter used in SPGL1 is set to $\eta\sqrt{m} = 0.06\|y_{\text{input}}\|_2$, which should yield a reconstruction error of $6\%$. Also, since the time points and the frequencies are at different scales, we multiply all learned frequencies by $\code{frqscale}$, which is set by default to $\code{frqscale} = \dfrac{T}{\omega_{\max}}$ before applying the clustering algorithm on the learned time-frequency pairs. The radius of a cluster's neighborhood in time-frequency space $\varepsilon$ is chosen based on the structure of the groups of modes in the spectrogram, noting that a larger value for $\varepsilon$ yields fewer clusters. When $\code{frqscale}$ is set to the default value, we used $\varepsilon = 0.2 T$ as a starting guess. Lastly, we set the number of modes in a neighborhood required to be considered a core point  $\code{min\_samples}=4\text{\ or\ } 5$.

In EWT, VMD, and SRMD, the number of intrinsic modes is specified for each experiment. For visualization and comparison, when the number of learned modes agrees or exceeds the number of true modes, we pair the learned and true modes based on the minimum $\ell_2$ distance. In particular, let $\{y_k^{\text{learned}}\}_{k=1}^{K}$ be the learned modes and $\{y_p^{\text{true}}\}_{p=1}^{P}$ be the true modes, then we pair the true and the learned modes by reindexing the learned modes using \begin{align*}
  y_{p}^{\text{learned-reindexed}} := \argmin_{k\in [K]} \|y_k^{\text{learned}} - y_p^{\text{true}}\|_2,
\end{align*}
for all $p\in [P]$. The remaining learned modes that are not paired are then combined into the learned mode with the highest error. The pairs $(y_{p}^{\text{learned-reindexed}}, y_{p}^{\text{true}})$, for $p\in [P]$, are used to compute the relative errors of our SRMD. In practice when the ground truth modes are not known, if there are $n$ more learned modes than true modes, we merge mode $n+1$ with the mode that has the smallest $\ell_2$ norm. The remaining modes are discarded. This is based on the assumption that the number of modes is known and their $\ell_2$ norms are comparable. For comparison with EMD, EEMD, and CEEMDAN, their extra modes are merged in the order they are extracted, i.e. the merging is based on their frequency scaling. This avoids the need to fine-tune their hyperparameters. This is based on the rationale that the excess modes occur from over-decomposition and that the mode with the highest error could be improved by merging the excess modes back together. Although the merging step may not be necessary for some applications, we use it to conduct a fair comparison on benchmark tests. Lastly, we plot the magnitude of the non-zero entries of the learned coefficient vector $c$ on the spectrogram. Note that all non-zero coefficients can be re-assigned to a positive value after shifting the phase of the corresponding basis term.  This leads to a sparse spectrogram representation of the signal. 

\subsection{Discontinuous Time-Series}\label{sec:discontinuity}
The first example is from \cite{daubechies2011synchrosqueezed}, where the input signal  $y(t) = y_1(t) + y_2(t) + y_3(t)$ for $t\in [0,2]$ is a composition of a linear trend $y_1(t)$, a pure harmonic signal $y_2(t)$, and a harmonic signal with a nonlinear instantaneous frequency $y_3(t)$:
\begin{equation}\label{eqn:discontinuity}
\begin{aligned}
    y_1(t) &= \pi t\, \text{1}_{[0,5/4)}(t)\\
    y_2(t) &= \cos(40\pi t) \, 1_{[0,5/4)}(t) \\
    y_3(t) &= \cos\left(\frac{4}{3} \left( (2\pi t - 10)^3 - (2\pi -10)^3\right) + 20\pi(t-1) \right)\, 1_{(1,2]}(t),
    \end{aligned}
\end{equation}
 and $1_{\mathcal{I}}(\cdot)$ denotes the indicator function over the interval $\mathcal{I}$.  The input signal has a sharp transition at $t= \frac{5}{4}$. The number of modes is fixed at $3$ for this experiment. The dataset contains $m=320$ points equally spaced in time from $[0,2]$ and the total number of random features in SRMD is set to $N=50m = 16000$. For DBSCAN, we set the minimum number of core points in a cluster to $min\_samples =3$ and the maximum distance between any two points in a neighborhood is set to $\varepsilon =0.1$.
\begin{figure}
    \centering
\includegraphics[width = 3.5  in]{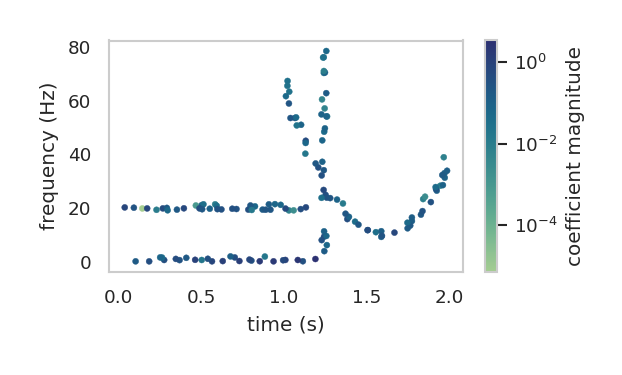}
\includegraphics[width = 2.9 in]{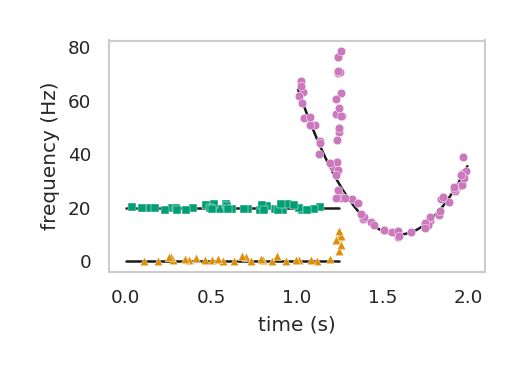}\\
\includegraphics[width = 6.25 in]{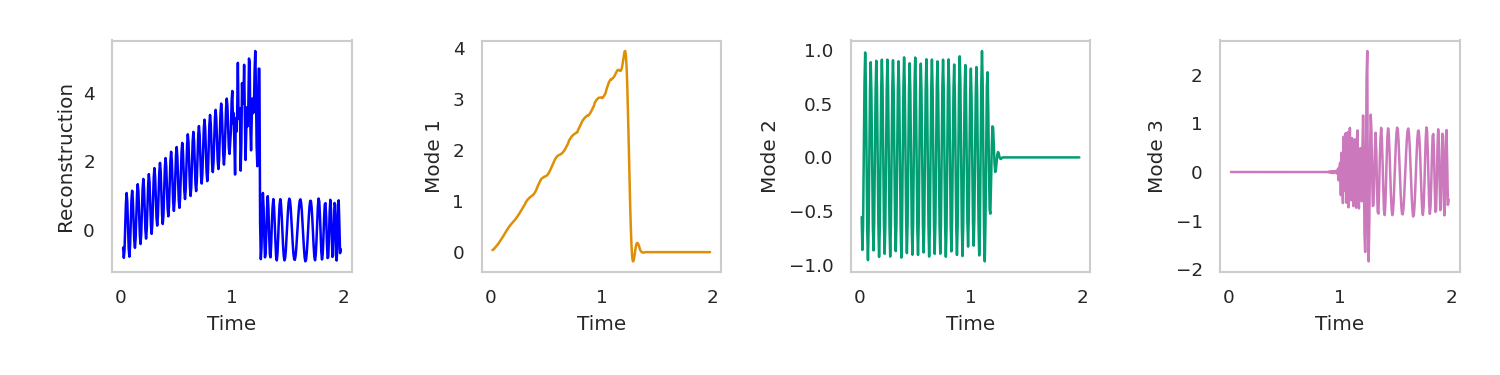}
\includegraphics[width = 6.25 in]{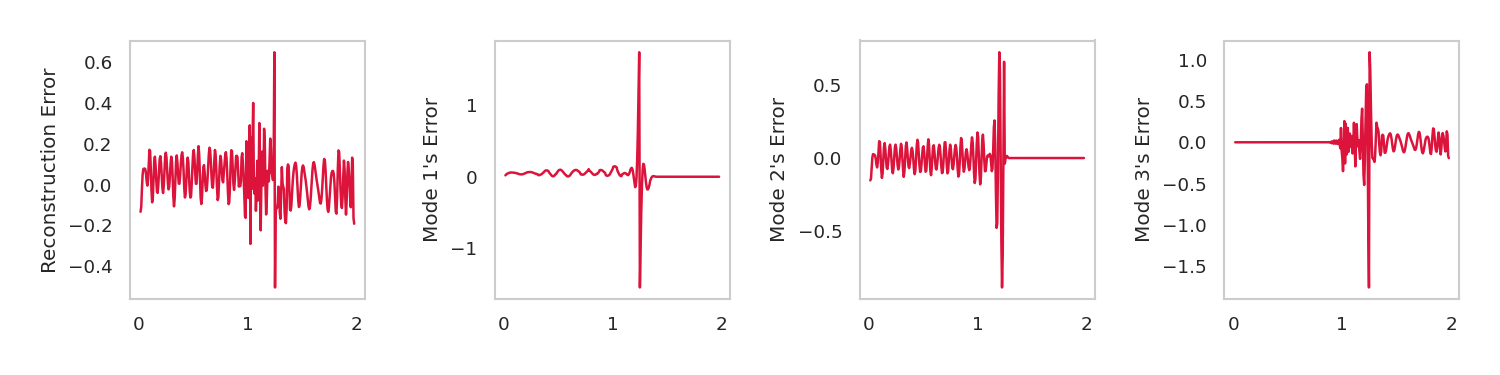}
    \caption{{\bf Example from Section \ref{sec:discontinuity}:} Top left: magnitude of non-zero learned  coefficients. Top right: clustering of non-zero coefficients into three modes.  Middle row from left to right: reconstructed signal (in blue) and the three extracted modes matching the colors of the top right clusters.  Last row: error of the reconstruction and the three modes compared to the ground truth.}
    \label{fig:ex_discontinuities_SWR_results}
\end{figure}

\begin{figure}[htp]
\centering
    \includegraphics[width=6.25 in]{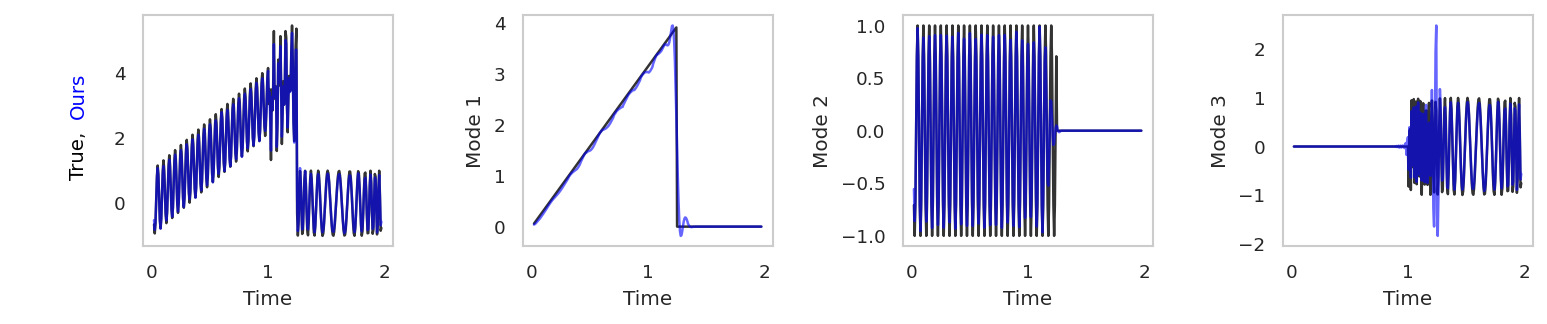}
    \includegraphics[width=6.25 in]{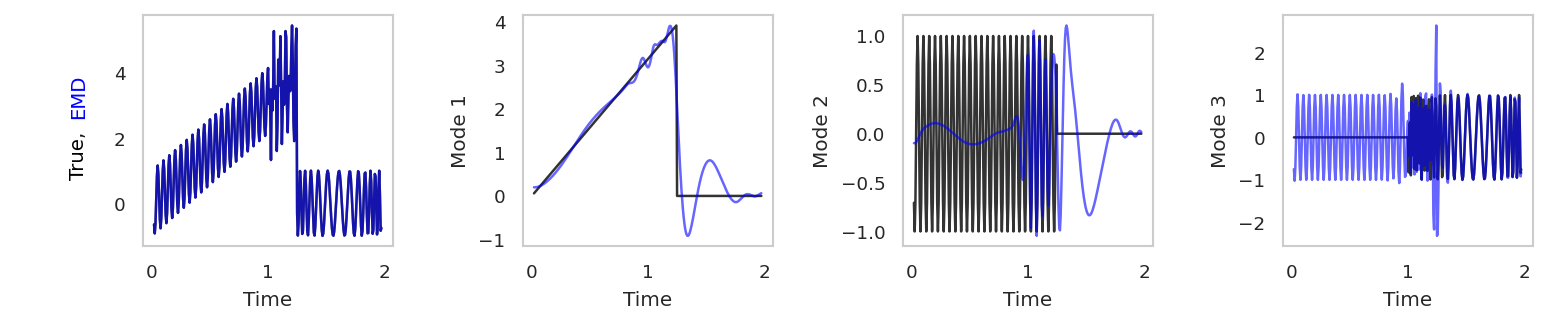}
    \includegraphics[width=6.25 in]{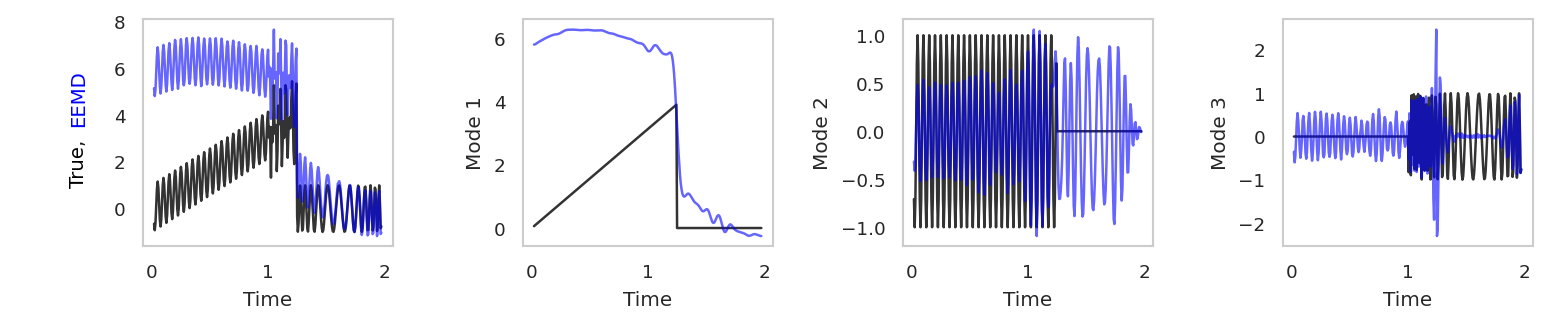}
    \includegraphics[width=6.25 in]{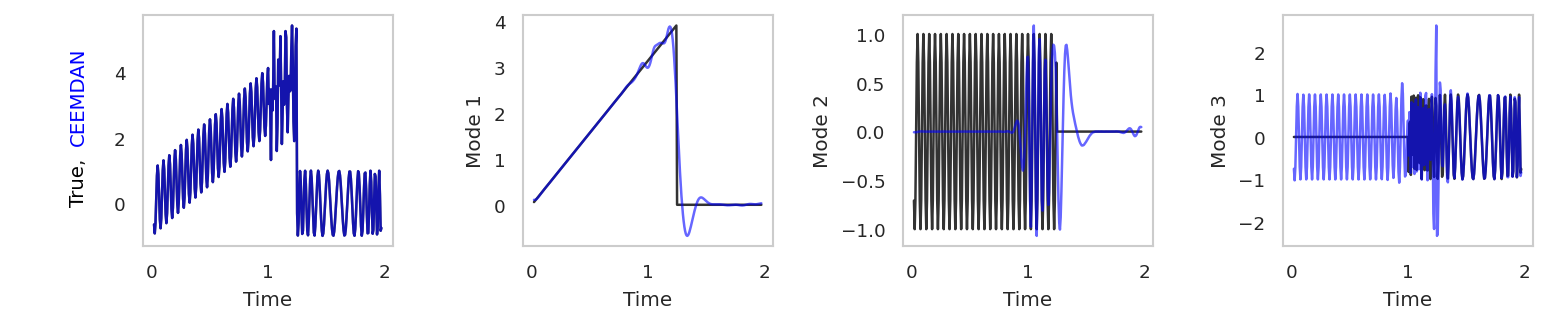}
    \includegraphics[width=6.25 in]{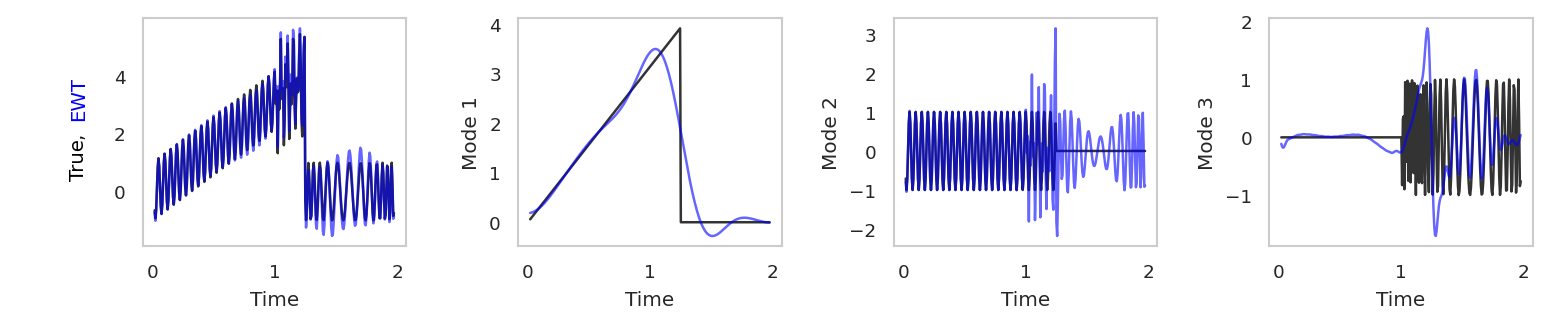}
    \includegraphics[width=6.25 in]{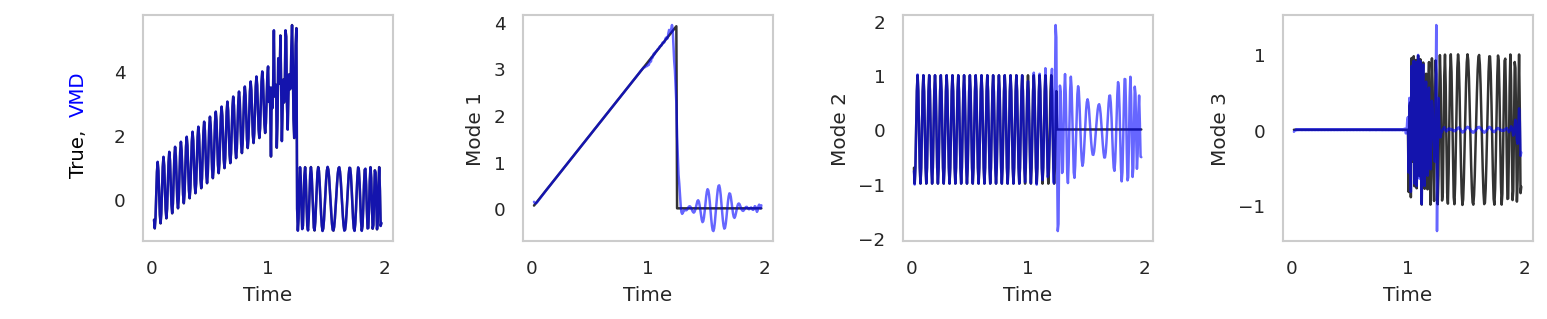}
    \caption{{\bf Example from Section \ref{sec:discontinuity}:} Comparing different methods on the discontinuous time-series example (Equation \eqref{eqn:discontinuity}). Top to bottom rows are our proposed method (SRMD), EMD, EEMD, CEEMDAN, EWT, and VMD. The first column shows the noiseless ground truth (in black) and the learned signal representation (in blue). The remaining three columns are the first, second, and third modes where the true IMFs are plotted in black and the learned IMFs are in blue.}    \label{fig:ex_discontinuities_all_comparison}
\end{figure}

Figure~\ref{fig:ex_discontinuities_SWR_results} shows that SRMD can decompose the discontinuous signal with minimal mode mixing, i.e.\ a clear separation between the three modes. In particular, the three learned SRMD modes are close to the ground truth modes with the errors mainly occurring at the point of discontinuity  $t=\frac{5}{4}$ (see also Figure~\ref{fig:ex_discontinuities_all_comparison}). It may be possible to avoid the error at the discontinuity with a choice of basis better suited to jumps such as the function associated with Haar wavelets (i.e.\ the square function) instead of the sine function. Even so, Figure~\ref{fig:ex_discontinuities_all_comparison} shows that the SRMD output using the sine function, whose wavelet equivalent is not well-suited to discontinuities, is still sufficient for our method to outperform existing methods. Moreover, the representation and clustering plots of Figure \ref{fig:ex_discontinuities_SWR_results} show that the sparse spectrogram and modal decomposition are indeed sharp and well-separated. This is a better separation than the various STFT and CWT, and their synchro-squeezed versions shown in Figures 1, 2, and 6 of \cite{daubechies2011synchrosqueezed}. Additionally, SRMD has the advantage of locating the individual regions that define each mode in the time-frequency domain. This could be done with a synchro-squeezed version of STFT or CWT by extracting a thin region around the instantaneous frequency curves; however, the reconstruction accuracy is not guaranteed. Specifically, using the synchro-squeezed approach, the selected regions may exclude areas of the time-frequency domain that contain important information, or double count intersecting regions like in Equation \eqref{eqn:intersecting}. 

In Figure~\ref{fig:ex_discontinuities_all_comparison}, the input signal and the three true intrinsic modes are shown in black, while the extracted modes from our and the other five methods are shown in blue. Comparing the errors, EMD and CEEMDAN reconstruct the signal with machine precision (the error is on the order of $10^{-16}$), VMD has a reconstruction error of only $0.4\%$, while EEMD and EWT have poor signal reconstruction. One possible reason for the errors in EEMD and EWT is that these methods do not guarantee that the sum of the reconstructed intrinsic modes equals the original signal. In terms of signal decomposition, the VMD and EWT extract the linear trend well (the second column of the last two rows in Figure~\ref{fig:ex_discontinuities_all_comparison}) while their learned second and third modes agree with the corresponding ground truth ones on the time interval $t\in [0,1]$ but create a false oscillatory pattern on the remaining interval $t\in [1,2]$. This indicates that the decomposition produced by the EWT and VMD approaches experiences a non-trivial amount of mode mixing. The remaining methods used in this comparison are unable to extract the true mode behaviors. The relative $\ell_2$-errors between the learned modes and the true modes and between the reconstructed signal and the noiseless ground truth signal of our SRMD, as well as various state-of-the-art intrinsic mode decomposition methods, is summarized in Table~\ref{tab:l2error}. Specifically, in Table~\ref{tab:l2error} we take the average of each mode over 100 trials to generate an ensemble version of SRMD (referred to E-SRMD), showing that the average over the modes is robust. 

\begin{table}[]
\centering
\begin{tabular}{|l|l|c|c|c|c|c|c|}
\hline
{Example} & Relative Error & E-SRMD & EMD & EEMD & CEEMDAN & EWT & VMD \\
\hline
\multirow{4}{*}{Section \ref{sec:discontinuity}} &  
Reconstruction & 6.4  & 0  & 172.5 & 0  & 9 & 0.4 \\ \cline{2-8} 
   & Mode 1 & {\color{blue}11.0} & 19.7   & 187.2 & 14.2 & 24.1 & 12 \\ \cline{2-8} 
   & Mode 2 & {\color{blue} 33.1}  & 109.9 & 77.4 & 101.1  & 88.3 & 70.5\\ \cline{2-8} 
   & Mode 3 & {\color{blue}38.4} & 111.1  & 93.3 & 111.2 & 119.3 & 86.7 \\ \hline
   \hline 
 \end{tabular}
 \caption{Relative $\ell_2$-errors (\%) between the reconstructed signal and the noiseless ground truth signal and between the learned modes and the true modes for the experiments detailed in Section \ref{sec:discontinuity}. For SRMD, we run an ensemble SRMD following the idea from EEMD, that is, we run SRMD 100 times and for each mode, we compute the error between the averaged learned mode and the true one. The smallest errors for each learned mode are highlighted in blue.} \label{tab:l2error}
\end{table}

\subsection{Instantanenous Frequencies of Intersecting Time-Series }\label{sec:intersecting}
For the second example, we use a challenging benchmark test from \cite{daubechies2011synchrosqueezed}: \begin{equation}\label{eqn:intersecting}
\begin{aligned}
    y_1(t) &= \cos\left(t^2 + t + \cos(t)\right), \\
    y_2(t) &= \cos(8t),
\end{aligned}
\end{equation}
where the true signal is given by
\[ y(t) = y_1(t) + y_2(t),\quad t\in [0,10].\]
The instantaneous frequencies of those two modes $y_1(t)$ and $y_2(t)$ (see the definition in \cite{daubechies2011synchrosqueezed}) are given by $\omega = (2t+1 -\sin t) $ and $\omega =  8$, respectively and thus intersect at about $t^*\approx 3.38$. This example shows our method's ability to obtain instantaneous frequencies and to deal with over-segmentation. For example, we expect large magnitudes of the coefficients to be near the instantaneous frequency curves and in phase.

The dataset contains $m= 1600$ equally spaced in time points from $[0,10]$ and we set $N = 10m = 16000$. Also, we set $\omega_{\max}=5$, since all frequencies are less than $5$\ Hz. Note that this scale is the physical frequency in $\Hz$ whereas the formula for the modes in this example are given in the angular frequency unit rad$/$s, thus we set \code{frqscale}$=2\pi$ rather than $1$. The maximum distance between any two points in a neighborhood in the DBSCAN clustering algorithm is set to $\varepsilon =2.0$.

\begin{figure}[htp]
    \centering
    \includegraphics[width=3.25 in]{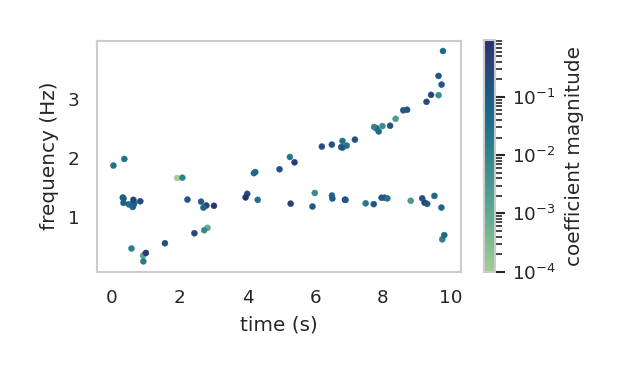}
    \includegraphics[width= 2.8 in]{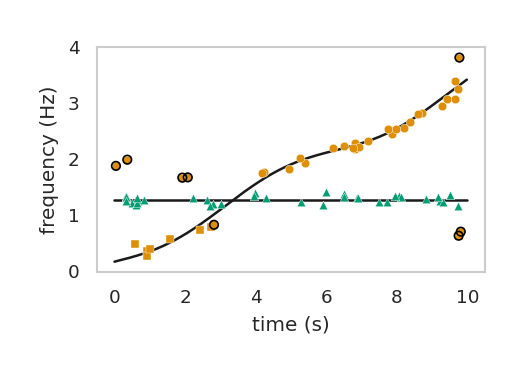}\\
    \includegraphics[width=2.9 in]{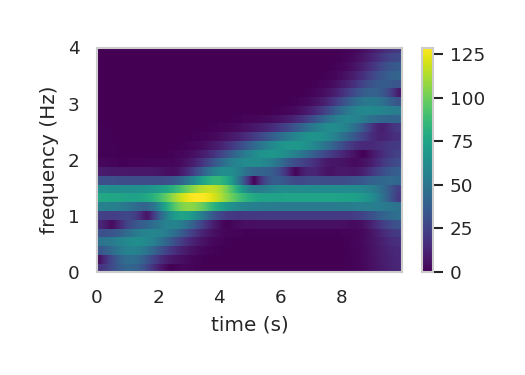}\quad\quad 
    \includegraphics[width=2.9 in]{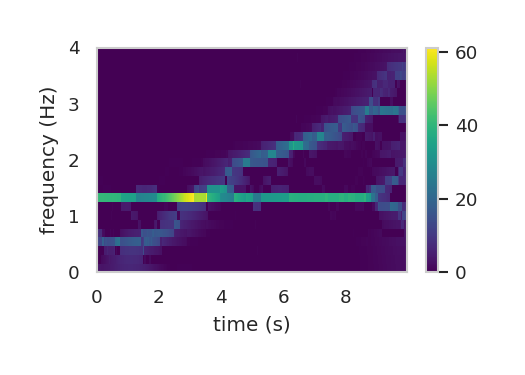}
    \includegraphics[width=3.15 in]{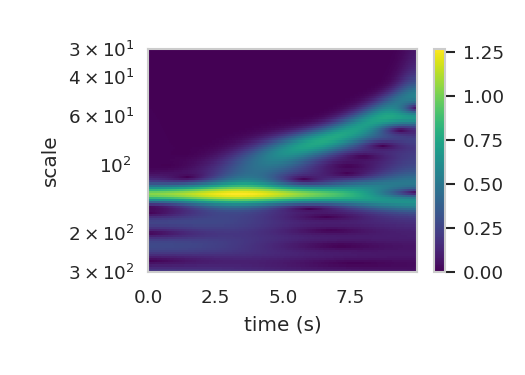}
    \includegraphics[width=3.15 in]{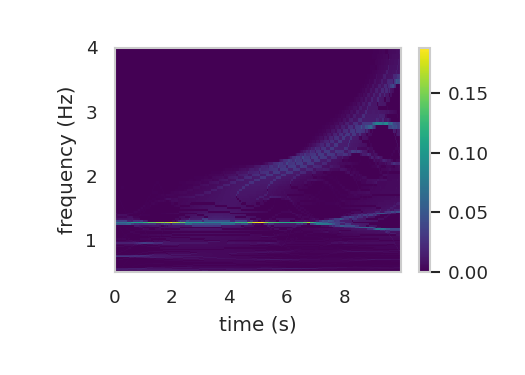}

    \caption{{\bf Example from Section \ref{sec:intersecting}:} First row displays the results of our method: a plot of the magnitude of the learned coefficients (top left) and a plot of the clusters (top right). Dots with a black outline indicate coefficients that DBSCAN labeled as noise and were re-labeled with the nearest cluster (in scaled frequency space). The true instantaneous frequencies of the two true IMFs are solid lines in black. The second row plots a zoom-in (frequency range $[0,4\,\text{Hz}]$) of the absolute values of the STFT and its synchrosqueezed version. The third row plots a zoom-in of the absolute values of the CWT and its synchrosqueezed version.}
    
    \label{fig:ex_intersecting_IF}
    
\end{figure}

The spectograms plot the pairs $\{(\tau_j,\omega_j)\}_j$ that are retained by the sparse optimization in SRMD, after discarding the pairs associated with zero coefficients, and seem to reveal the instantaneous frequencies of the full input signal as indicated in the first row of Figure \ref{fig:ex_intersecting_IF}. Moreover, our method obtains a clearer spectrogram than the STFT, CWT, synchrosqueezed transforms based on wavelets (Figure 8 in \cite{daubechies2011synchrosqueezed}), modified STFT \cite{gaurav2011synchroSTFT}, and CWT \cite{thakur2013synchrosqueezing} (see Figure \ref{fig:ex_intersecting_IF}). For the STFT and its synchrosqueezed results (second row), a Gaussian window with standard deviation $0.75\ \text{seconds} = 60\ \text{samples}$, Fourier transform width of $512$ samples, and hop-size of 1 are used. The standard deviation was chosen to match the window size used in our method. For the CWT and its synchrosqueezed version (third row), 232 scales were chosen between 3.8 samples ($\approx{}21$\,Hz) and 512 samples ($\approx{}0.16$\,Hz) with logarithmic spacing (the default settings  of the package \code{ssqueezepy}, see also \cite{muradeli2020ssqueezepy}), and so the maximum scale matches the Fourier transform width in the STFT used. Recall that the formula used to convert between scales and frequencies is: $\text{frequency} = \frac{\text{sample rate}}{\text{scale}}$. To have a fair comparison, the Morlet wavelet was used since it utilizes a Gaussian window. In the Appendix, we display the results of merging modes, when the number of learned modes obtained from DBSCAN is greater than the number of true modes (see Figure \ref{fig:ex_intersecting_3modes} and Figure \ref{fig:ex_intersecting_2modes} in Section \ref{sec:appendixIntersecting}). More comparisons on the intersection time series signal are also shown in the Appendix (see Figure \ref{fig:ex_intersecting_all_comparison} in Section \ref{sec:appendixIntersecting}).

\subsection{Pure Sinusoidal Signals with Noise}\label{sec:VMD_eq32}
In this example, we decompose a noisy signal into three modes where the noise level has a larger amplitude than one out of the three modes. The input signal is defined as \cite{dragomiretskiy2013variational}:
\begin{align}
    y(t) = \cos(4\pi t) +\frac{1}{4}\cos(48\pi t) +\frac{1}{16}\cos(576\pi t) + \varepsilon, \quad \varepsilon\sim \mathcal{N}\left(0, 0.1\right).
\end{align}
The hyperparameters are set to $\omega_{max} = 500$, $\Delta=2s$, $m=1000$, $N=50m$, $r=15\%$, $\code{threshold}=0$, $\varepsilon=1.5$ (for DBSCAN), $\code{min\_samples}=4$, and $\code{frqscale}=1$. Before training, the signal is extended from the domain $[0,1]$ to $[-1,2]$ by an even periodic extension. The number of data points $m$ and features $N$ are thus scaled by a factor of $3$ during the training phase. To verify the stability of SRMD with respect to the randomness of the random feature matrix, we run SRMD 100 times. For each time, we compute the relative error between the noiseless ground truth and the reconstructed signal as well as  the relative errors between the true modes and the corresponding learned modes, under the assumption that the number of learned modes is given (which is three for this example). We report $(\mathtt{mean},\mathtt{std})$ of SRMD's relative errors as well as the relative errors of benchmark methods in Table \ref{tab:l2error2}. More tests for this example are shown in the Appendix (see Section \ref{sec:appendixPureSinusoidal}). Our methods provide the smallest relative errors of the extracted modes for both cases.

\begin{table}[]
\centering
\begin{tabular}{|l|l|c|c|c|c|c|c|}
\hline
{Example} & Relative Error & SRMD & EMD & EEMD & CEEMDAN & EWT & VMD \\
\hline
       \multirow{4}{*}{Section \ref{sec:VMD_eq32}} & 
  Reconstruction & (1.6,0.6) &14.1  & 14.3   & 14.1  & 13.6 &5.3  \\ \cline{2-8} 
   & Mode 1 & $(1.0,0.6)$   &  25.6 & 25.4 &    25.8 & 2.5  & 2.3 \\ \cline{2-8} 
   & Mode 2 & $(3.2,1.3)$  &  112.5  & 110.8 &   111.8& 59.1 & 12.1 \\ \cline{2-8} 
   & Mode 3 & $(13.6,6.5)$  &  148.1  & 122.2 &   143.8& 100.5 & 53.9 \\ \hline 
    \hline
    \hline
  \end{tabular}
 \caption{Relative $\ell_2$-errors (\%) between the reconstructed signal and the noiseless ground truth signal and between the true modes and the corresponding learned modes for experiments in Section \ref{sec:VMD_eq32}. For SRMD, we run 100 times and for each time, we compute the relative errors between the learned modes and the true ones. The results in the SRMD column are $(\mathtt{mean},\mathtt{std})$ of the relative errors. } \label{tab:l2error2}
\end{table}

\subsection{Musical Example}\label{sec:music}
We use our method to decompose a two-second clip of a guitar and flute playing simultaneously. The representation into sparse random features is performed identically as before, but the clustering is performed by splitting spectrogram with a frequency above and below the frequency cutoff of 480\Hz. This cutoff is chosen from the visual information provided by the plot of nonzero random features' time-shift and frequency. This is similar to traditional signal processing techniques that rely on STFT to observe and isolate regions in time-frequency space. Our method has the advantage of finding a sparse representation so individual harmonics are better defined and the signal is denoised in the process.

\begin{figure}[htp]
    \centering
    \includegraphics[width=\linewidth]{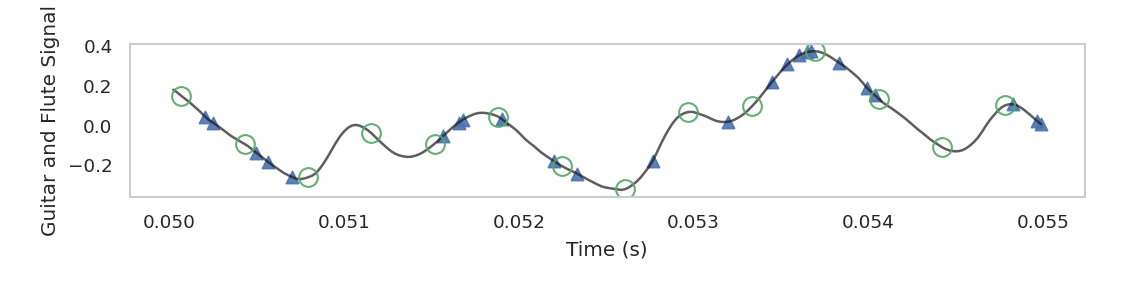}
     \caption{ {\bf Example from Section \ref{sec:music}:} Zoomed in plots. Original signal is sampled at 44.1kHz, equally-spaced downsampling is at 2.8kHz (circles), and random downsampling (solid triangles) by a factor of 16.}
    \label{fig:ex_music_data}
\end{figure}

We examine the decomposition results of our method in two cases when the input signal is either equally-spaced downsampling or random downsampling. An illustration of the sampled data in both cases is presented in Figure \ref{fig:ex_music_data}, where the original full signal is sampled at 44.1k\Hz, the equally-spaced downsampling is at 2.8k\Hz, and the random sampling has $1/16th$ as many points as the original full signal. The hyperparameters of our method are $m = 5107$, $N = 10m$, $\Delta = 0.03$,  and $r = 10\%$. For the equally spaced downsampling, we set $\omega_{\max} = 2.8k\Hz / 2 = 1378\Hz$ (the Nyquist frequency), whereas the random sampling uses twice this with $\omega_{\max} = 2.8k\Hz$.

\begin{figure}[htp]
    \centering
    \minipage{\linewidth}
    \includegraphics[width=0.49\linewidth]{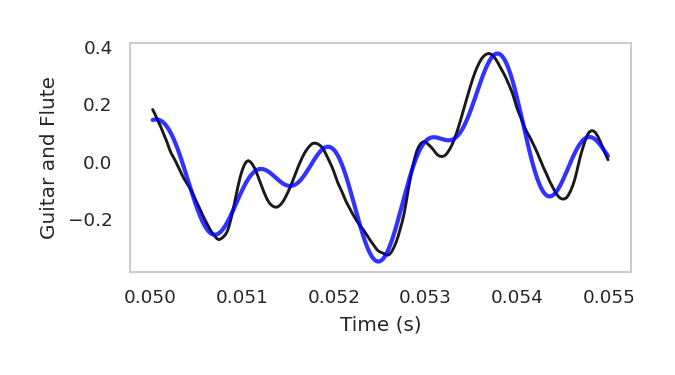}
    \includegraphics[width=0.49\linewidth]{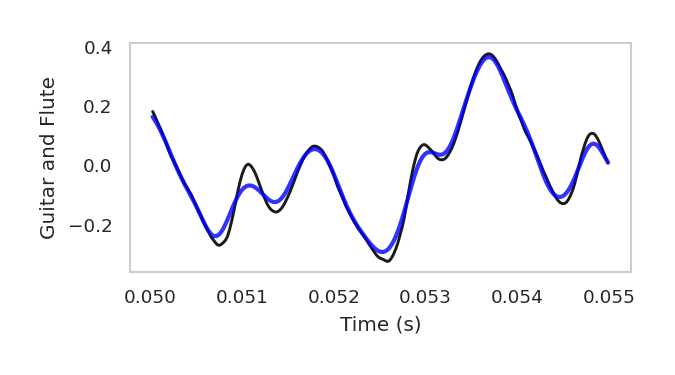}
    \endminipage
    \\
    \minipage{\linewidth}
    \includegraphics[width=0.49\linewidth]{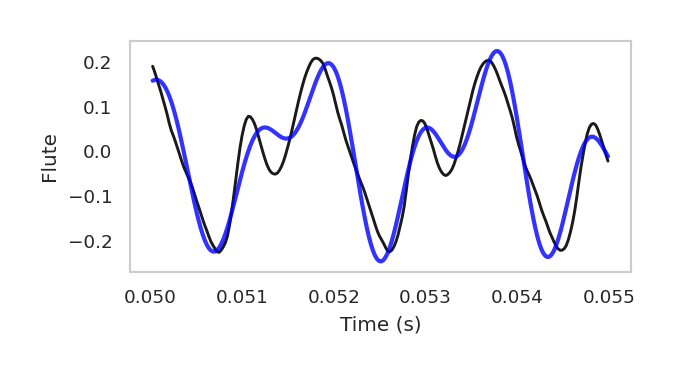}
    \includegraphics[width=0.49\linewidth]{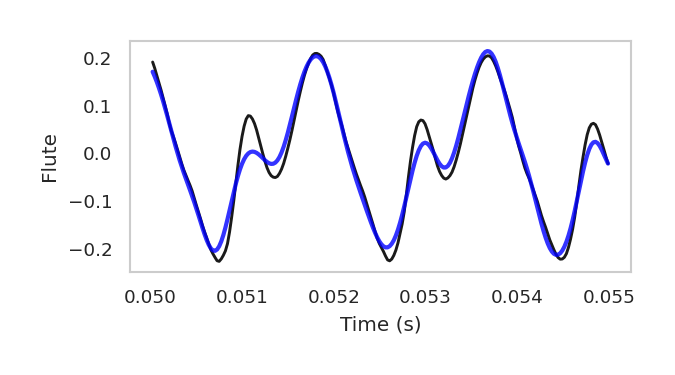}
    \endminipage
    \\
    \minipage{\linewidth}
    \includegraphics[width=0.49\linewidth]{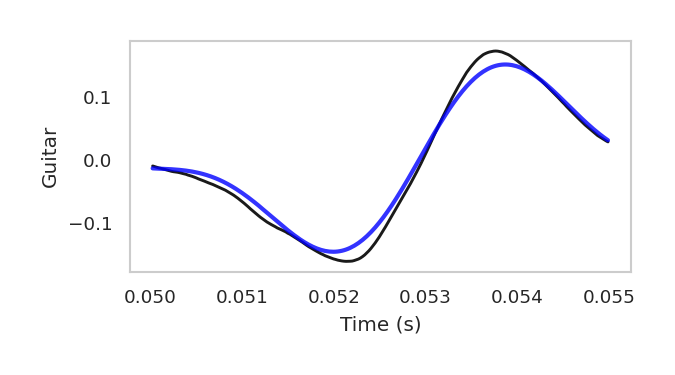}
    \includegraphics[width=0.49\linewidth]{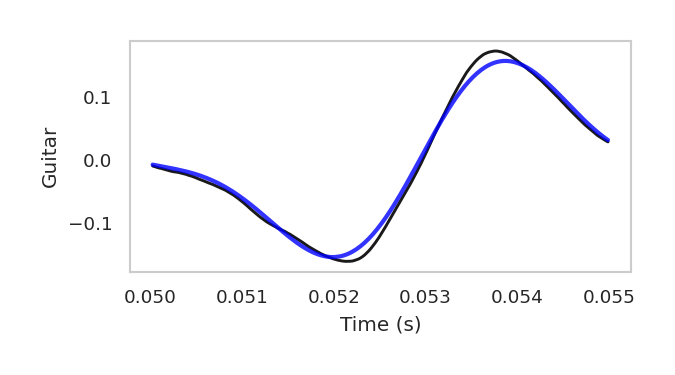}
    \endminipage
    
    \caption{{\bf Example from Section \ref{sec:music}:}
    Zoomed-in plots of the learned signals and modes (blue) using SRMD as compared to the original fully sampled signal (black). Left: results using equally-spaced data. Right: results using random samples. The input signal is downsampled by a factor of 16.} 
    \label{fig:ex_music_equalvsrandom}
\end{figure}

The zoomed-in plots of the learned signals and modes from the equally-spaced and random downsampled data (by a factor of 16) are plotted in Figure \ref{fig:ex_music_equalvsrandom}. The results indicate that our method works well even when the given data is sampled randomly. Moreover, random sampling allows for higher frequencies in the flute to be captured since we are not limited by the typical Nyquist rate. In this case, the maximum frequency of the features generated is twice as high as the equally sampled, while maintaining the same number of generated features.

\begin{figure}
    \centering
    \minipage{\linewidth}
        \includegraphics[width=0.55\linewidth]{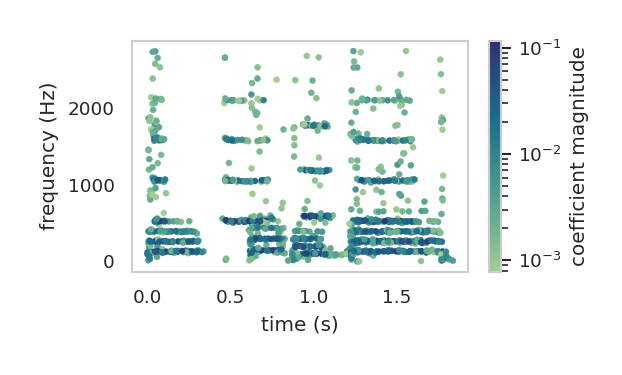}
        \includegraphics[width=0.45\linewidth]{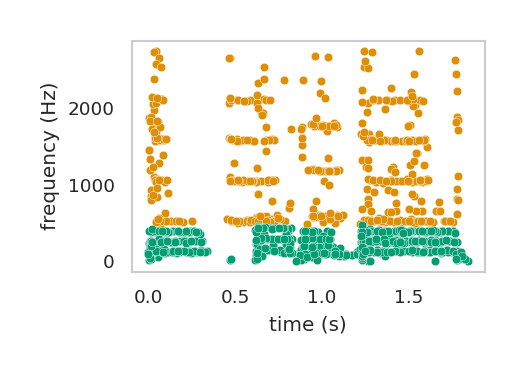}
    \endminipage\\
    \minipage{\linewidth}
        \includegraphics[width=\linewidth]{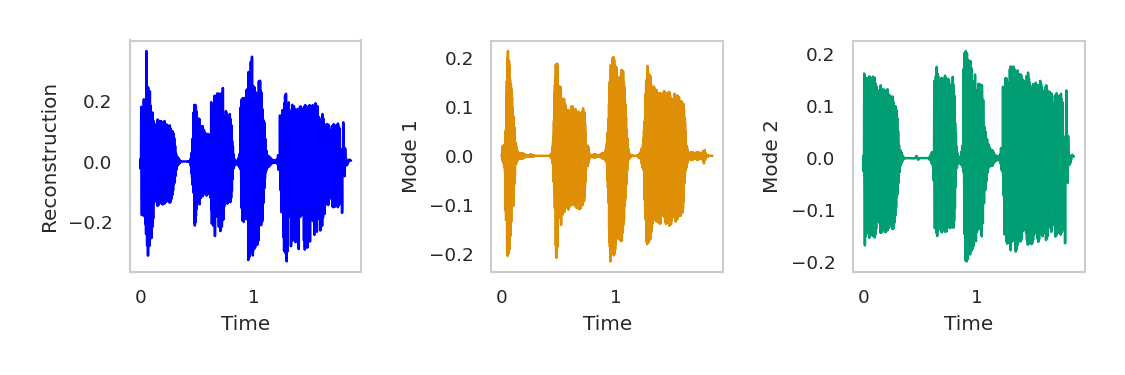}
        \includegraphics[width=\linewidth]{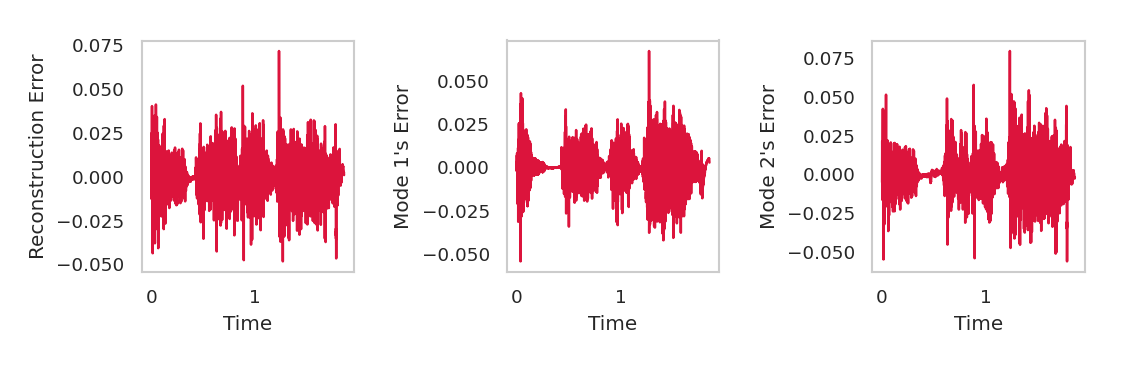}
    \endminipage
    \caption{{\bf Example from Section \ref{sec:music}:} Results of our method with randomly downsampled input by a factor of 16. First row: magnitude of the non-zero learned coefficients (left), two learned clusters (right). Second row: Learned signal (blue, left) and the learned flute and guitar modes (middle and right). Last Row: error between the learned signals and the downsampled input. }
    \label{fig:ex_music_SWR_results}
\end{figure}

In Figure~\ref{fig:ex_music_SWR_results}, we plot the magnitude of learned non-zero coefficients (top left) and the clustering of those coefficients into two modes (top right), the reconstruction signal and the two learned modes (second row), as well as the corresponding errors (third row). The learned audio files (stored on GitHub \footnote{\url{https://github.com/GiangTTran/SparseRandomModeDecomposition}} are close to the ground truth.

\subsection{Gravitational Data}\label{sec:astronomy}
\begin{figure}[htp]
    \centering
    \minipage{0.49\textwidth}
         \includegraphics[width = \linewidth]{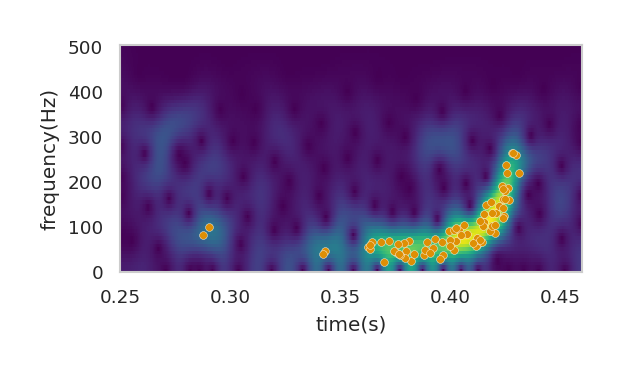}
        
        \includegraphics[width =\linewidth]{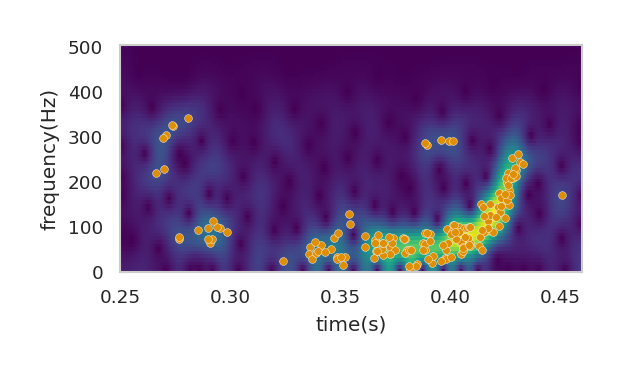}
    \endminipage
    \minipage{0.49\textwidth}
        \includegraphics[width=\linewidth]{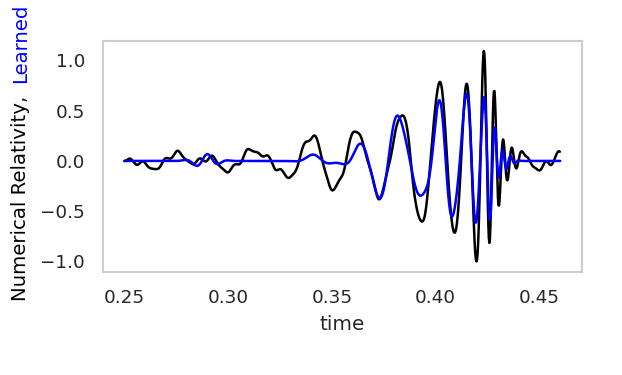}
        \includegraphics[width=\linewidth]{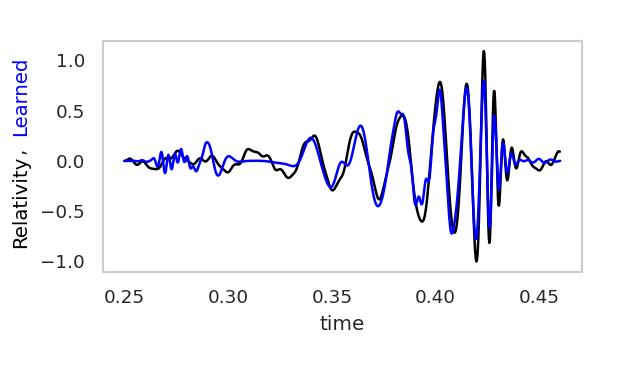}
    \endminipage
    \caption{{\bf Example from Section \ref{sec:astronomy}:} Left column: The largest 3\% (first row) and 5\% (second row) learned nonzero coefficients overlayed on an STFT in the time-frequency domain. Right column: Corresponding learned signals (blue) and numerical relativity data (black). Time is given as seconds after September 14, 2015, at 09:50:45 UTC. }
    \label{fig:astronomical}
\end{figure}

For a data-assisted discovery problem, we apply our method to a space-time distortion dataset which was the first to observe gravitational waves \cite{abbott2016}. The noisy input data is preprocessed in the same way as in \cite{abbott2016}, which consists of whitening, filtering, and downsampling steps. The preprocessing steps are necessary in order to enhance the nonlinear wave structure hidden under a layer of biases and noise that cannot immediately be seen from the data directly outputted by the instruments. In addition, we also normalize the input data by dividing the signal by the maximum value in order since the original signal is on the order of $10^{-19}$. The goal is to obtain an approximation to the waveform that captures the merging event and to denoise the signal. The hyperparameters for our method are set to $m = 861$, $N = 20 m$, $\Delta = 0.01$, and  $\omega_{\max} = 2048\Hz$. The parameter for the LASSO algorithm is {\color{magenta}$\lambda  = 12$.}  The top 3\% or 5\% largest non-zero coefficients show the merging and ringdown events in Figure~\ref{fig:astronomical}. For comparison, we show the numerical relativity curves, i.e.\ the space-time distortion measure on the gravitational interferometer.

\subsection{Parameter Tuning and Limitations}\label{sec:parameters}
We have shown that SRMD can better capture the correct modal separation, but for a full comparison, we discuss one of the potential difficulties over the other mode decomposition and signal representation approaches. In particular, the method has several parameters that the user needs to tune for the particular problem.
The tunable hyperparameters are the random feature matrix parameters $(\omega_{\max},\Delta,N)$, the optimization parameter parameter $\eta$ for the SPGL1 algorithm (or $\lambda$ in LASSO), and DBSCAN's parameters $\code{min\_samples}$, and $\varepsilon$. The frequency is rescaled by $\code{frqscale} = \dfrac{T}{\omega_{\max}}$ which does not need tuning. This value rescales the frequency-time domain to improve the performance of the clustering algorithm. As a general rule-of-thumb, we set the maximum $\omega$ scale to $\omega_{\max} = \dfrac{m}{2T}$ which can be tuned but was not necessary on the experiments in this work. 
The main ``free'' hyperparameters are $N$, $\eta$, $\Delta$, and $\varepsilon$. In the experiments, the number of randomly generated features $N$ varied between $5m$ to $50m$, based on \cite{chen2021conditioning, hashemi2021generalization}. In practice, while one could perform a coarse-scale search, i.e. running the algorithm with a set of possible $N$ in the range provided and evaluating the solution, the results are consistent within a large range of $N$ due to the sparse solver. 
For the basis pursuit parameter (in SPGL1), we used $\eta\sqrt{m} = 0.06\|y_{\text{input}}\|_2$, which seem to consistently yield a reconstruction error of $6\%$. We must allow for some error so that we can obtain a sparse representation, i.e. balancing regularity and representation, but the choice of relative error is data-dependent.  \\
The sensitive tunable parameters are the window size $\Delta$ and the clustering algorithm. The window size is set to $\Delta=0.1$ as default but does require some trial-and-error to match the complexity of the signal's frequency spread. The clustering neighborhood scale $\varepsilon$ and the core points $\code{min\_samples}$ also require user tuning. While we started with $\varepsilon = 0.2 T$ as an initial guess, the clustering component of the algorithm can have a small usable parameter range. {\color{teal}In the appendix, we provide two examples where the output may not be reasonable when the hyperparameters $\Delta$ or the DBSCAN hyperparameter $\varepsilon$ are not sufficiently tuned.} \\
Some additional limitations of this approach are as follows. If the random sampling of the time-frequency domain does not properly cover the true frequencies, then the method will perform poorly. This could be the case when one uses a very small number of features, but it is often avoided by the suggested parameter discussion above.  The choice of $\omega_{\max}$ is to match the Nyquist sampling rate, which can be lowered if one has prior information on the maximum frequency in the data. This can be an issue in some datasets, but when prior information is limited then a large time-frequency range would result in an increase in the computational cost. When the time-frequency domain has clustered sparse regions, SRMD is robust since we leverage the sparse random feature approach.

\section{Summary}
We proposed a random feature method for approximating signals using a sparse time-frequency representation. The model can be used for signal representation, denoising (including outlier or corruption removal), and mode decomposition. The sparsification of the spectrogram leads to a clearer separation between clusters and thus an adaptive mode decomposition. Compared to other state-of-the-art mode decomposition approaches, the SRMD is able to mitigate the issue of mode mixing, provide better separation between intersecting or overlapping spectrograms, and can capture jumps with fewer Gibbs artifacts. In our experiments, we showed that the active (non-zero entries) of the sparse random feature coefficients concentration around the instantaneous frequencies and thus can provide additional physical information for use in diagnostics and data-driven analysis. One important distinction from other approaches is that SRMD is not dependent on the sampling process, thus data can be obtained in a random or non-uniform fashion. In addition, one can extend the coherence-based results  \cite{hashemi2021generalization} or the restricted isometry property for random features \cite{chen2021conditioning} to show that the learned model using the SRMD algorithm has a small generalization error. 

\section*{Acknowledgement}
N.R. and G.T. were supported in part by NSERC RGPIN 50503-10842. H.S. was supported in part by AFOSR MURI FA9550-21-1-0084 and NSF DMS-1752116. The authors would also like to thank Rachel Ward for valuable discussions.

\bibliographystyle{plain}

\section{Appendix}
As additional examples, we include more comparison results on the intersection time series signal described in Section \ref{sec:intersecting} as well as the experiments on a noisy signal where the noise level has a larger amplitude than one of the modes (see Section \ref{sec:VMD_eq32}) and on an overlapping and noisy signal (see Section \ref{sec:overlapping}).
\subsection{Comparing Different Methods on the Intersecting Time Series Example}\label{sec:appendixIntersecting}
\begin{figure}[htp]
    \centering
    \includegraphics[width=\linewidth]{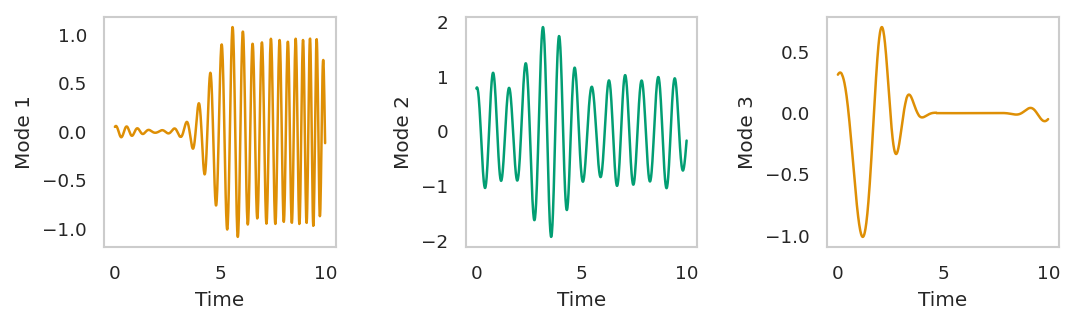}
       \caption{{\bf Example from Section \ref{sec:intersecting}:} The three modes from SRMD associated with the clusters: orange circles, green triangles, and orange squares from Figure \ref{fig:ex_intersecting_IF} (left to right).}
    \label{fig:ex_intersecting_3modes}
\end{figure}

\begin{figure}[htp]
    \centering
    \includegraphics[width=\linewidth]{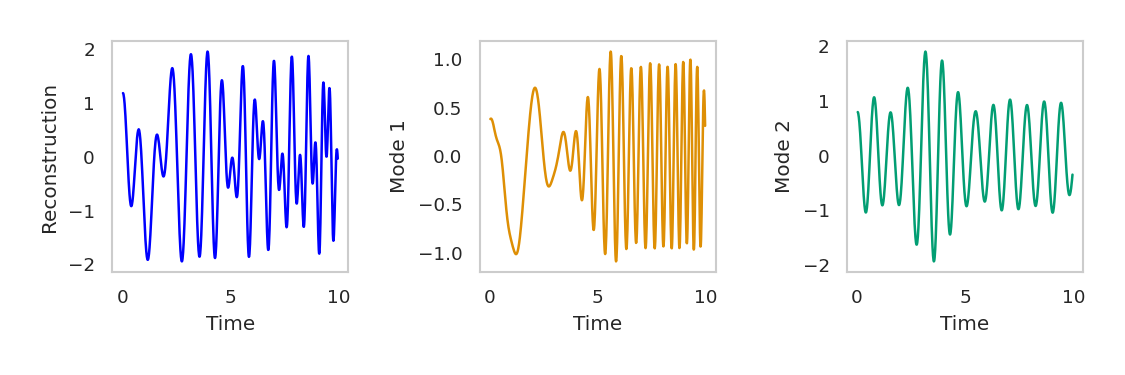}
        \includegraphics[width=\linewidth]{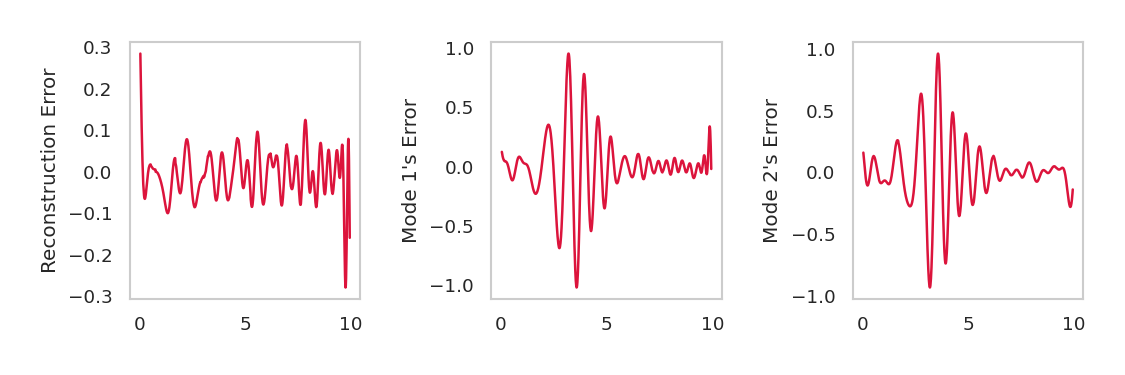}
\caption{{\bf Example from Section \ref{sec:intersecting}:} Decomposition results of our proposed SRMD method into two modes. First row: noiseless ground truth signal (in black) with the learned signal (in blue) of the full signal (top) and the two learned modes. Last row from left to right: Errors between noiseless ground truth and the learned representation, between the true modes and the extracted modes.}
    \label{fig:ex_intersecting_2modes}
\end{figure}

\begin{figure}[htp]
    \centering
    \includegraphics[width=3.9 in]{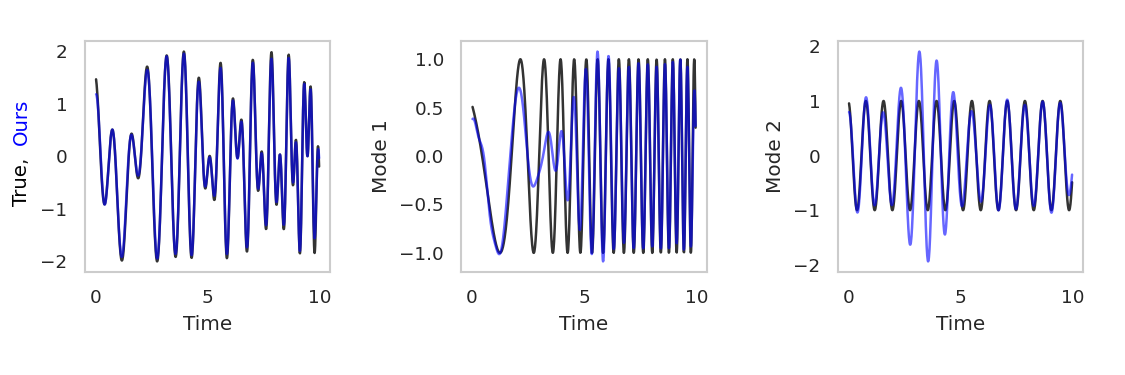}
    \includegraphics[width=3.9 in]{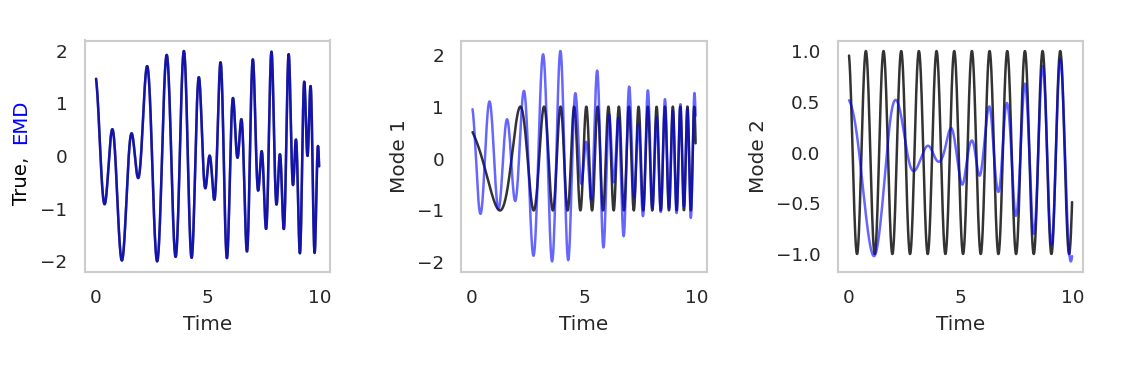}
    \includegraphics[width=3.9 in]{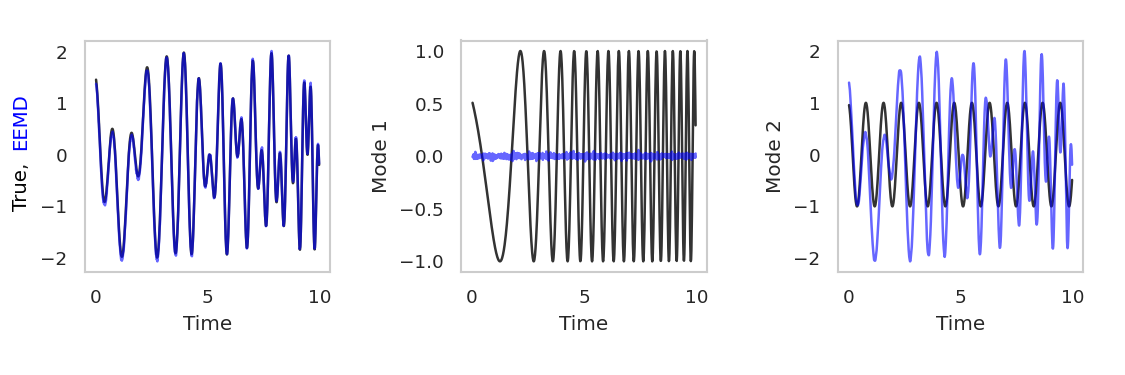}
    \includegraphics[width=3.9 in ]{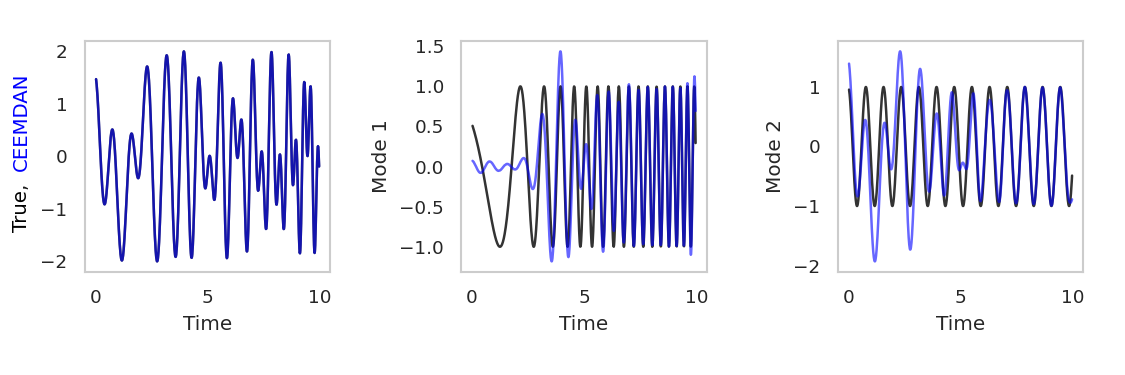}
    \includegraphics[width=3.9 in]{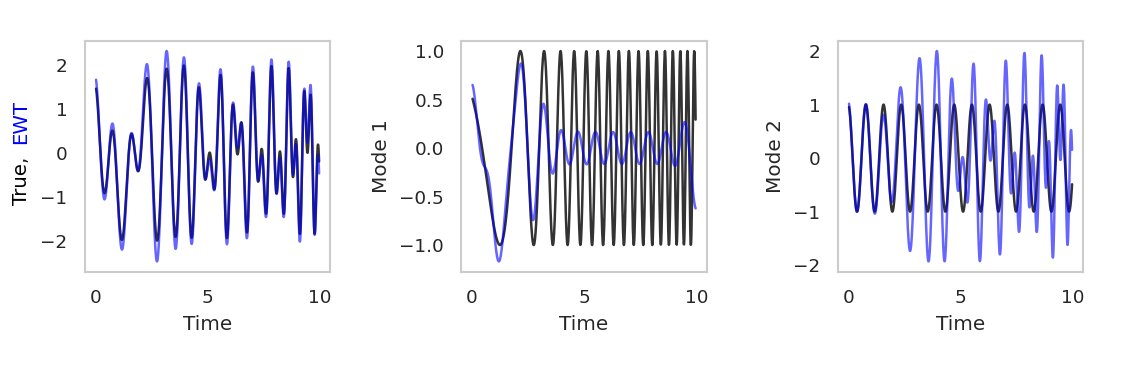}
    \includegraphics[width=3.9 in]{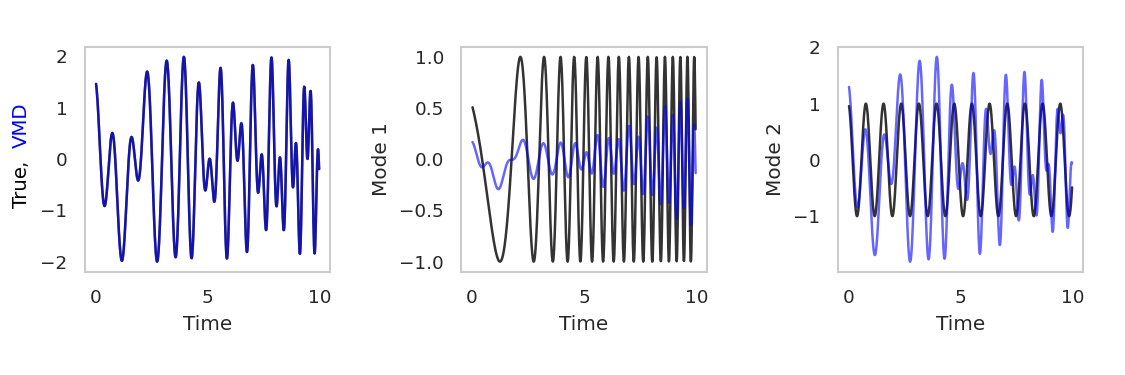}
    \caption{{\bf Example from Section \ref{sec:intersecting}:} Comparing different methods on the intersecting time-series example. Top to bottom rows are SRMD, EMD, EEMD, CEEMDAN, EWT, and VMD. The first column displays the noiseless ground truth (in black) and the learned signal representation (in blue). The remaining two columns are the two modes, where the true IMFs are plotted in black and the learned IMFs are in blue.}
    \label{fig:ex_intersecting_all_comparison}
\end{figure}

In this section, we present our reconstructed signals as well as our learned modes from the challenging intersecting time series in Section \ref{sec:intersecting}. Specifically, applying the DBSCAN on the extracted time-frequency pairs $\{(\tau_j,\omega_j)\}_j$ yields three clusters denoted by green triangles, orange circles, and orange squares (see Figure \ref{fig:ex_intersecting_IF} (right figure in the first row)). The corresponding learned modes are plotted in Figure~\ref{fig:ex_intersecting_3modes}. To reduce the decomposition to two modes, we keep the mode with the largest $\ell_2$ norm and combine the two learned modes with the smallest $\ell_2$-norm to construct the second mode, this is shown in Figure~\ref{fig:ex_intersecting_2modes}. Our proposed algorithm provides a reasonable extraction of modes where the errors between the learned modes and the true ones are almost zero everywhere, except on a time-shift region corresponding to the intersection of instantaneous frequencies. As seen in Figure~\ref{fig:ex_intersecting_all_comparison}, the other approaches have difficulty obtaining the two modes, likely due to the intersecting of frequencies in the spectrogram.  

In Figure \ref{fig:ex_intersecting_all_comparison}, we compare the reconstruction and the decomposition results of our method versus those obtained from some of the state-of-the-art intrinsic mode decomposition methods (EMD, EEMD, CEEMDAN, EWT, and VMD) on the intersecting time series signal given in Section \ref{sec:intersecting}. 
\subsection{Comparison Results on Pure Sinusoidal Signals with Noise} \label{sec:appendixPureSinusoidal}

\begin{figure}[htp]
    \centering
\includegraphics[width = 3.5  in]{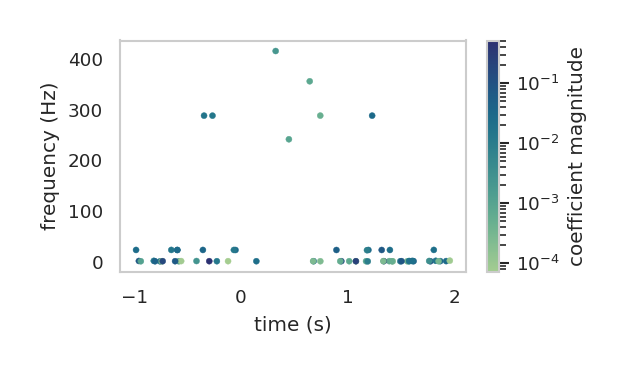}
\includegraphics[width = 2.9 in]{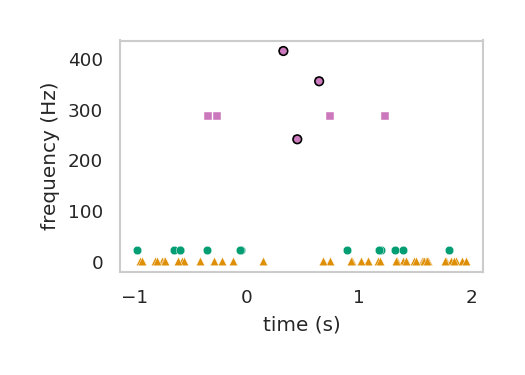}\\
\includegraphics[width = 6.25 in]{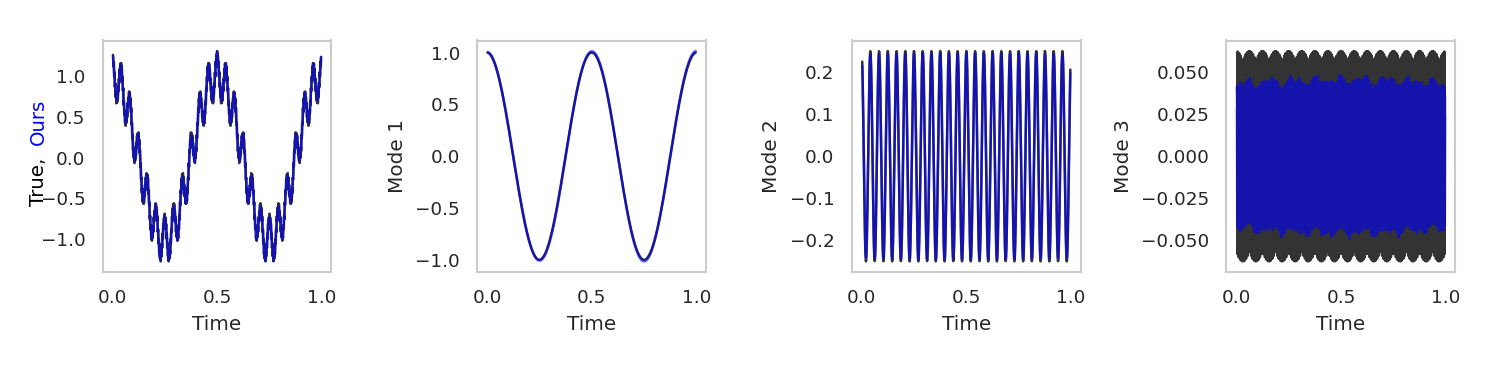}
\includegraphics[width = 6.25 in]{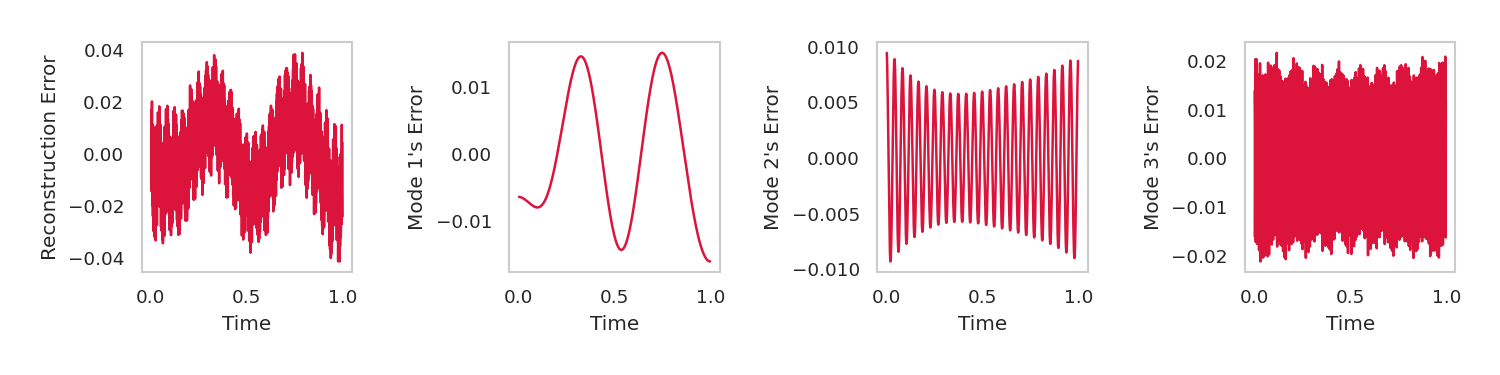}
    \caption{{\bf Example from Section \ref{sec:VMD_eq32}:} First row: The noisy input signal. Second row: Magnitude of non-zero learned coefficients (left) and learned clusters (right).  Third row from left to right: reconstructed signal (in blue) and the three extracted modes (in blue) versus the corresponding noiseless signal and modes (in black).  Last row: error of the reconstruction and the three IMFs compared to the ground truth.}
    \label{fig:ex_VMD_eq32_SWR_results}
\end{figure}

In this section, we present our reconstructed signals as well as our learned modes from the challenging noisy tri-harmonic signal described in Section \ref{sec:VMD_eq32}. In particular, in Figure \ref{fig:ex_VMD_eq32_SWR_results}, we plot the time-frequency pairs associated with non-zero coefficients (top left), the clustering of those non-zero coefficients (top right), the reconstruction signal and the three learned modes (second row), and the corresponding errors (third row). Our method can extract the first two learned modes with high accuracy. Note that both VMD (see \cite{dragomiretskiy2013variational}) and our method (see Figure \ref{fig:ex_VMD_eq32_SWR_results}) have difficulty in extracting the weak and high-frequency mode $y_3(t) = \frac{1}{16}\cos(576\pi t)$. Nevertheless, our method can identify the frequencies of all three modes. More precisely, the median frequencies of the three learned clusters are 1.99, 24.03, and 288.02\Hz, which is very close to the ground truth frequencies 2, 24, and 288\Hz.

\subsection{Overlapping Time-Series with Noise}\label{sec:overlapping}
In this experiment, we investigate an example with overlap. The input signal $y(t)$ is the summation of two modes $y_1(t) = \ifourier{Y_1}(t)$ and $y_2(t) = \ifourier{Y_2}(t)$ with overlapping frequencies and is contaminated by noise:
\begin{equation}\label{eqn:overlapping}
y(t) = y_{true}(t) + \varepsilon = y_1(t) + y_2(t) + \varepsilon,\quad \varepsilon\sim \mathcal{N}\left(0, \dfrac{r \|y_{true}\|_2}{\sqrt{m}}\right).  
\end{equation}
Here $\ifourier{Y_i}$ denotes the inverse Fourier transform of $Y_i$ for $i=1,2$, where
\begin{equation}
\begin{aligned}
   Y_1(k) = me^{-i\pi k} \left (e^{-\frac{9(k-16)^2}{32}}-e^{-\frac{9(k+16)^2}{32}}\right),     Y_2(k) = me^{-i\pi k} \left(e^{-\frac{9(k-20)^2}{32}}-e^{-\frac{9(k+20)^2}{32}}\right),
\end{aligned}
\end{equation}
for $k\in \mathbb{Z}$ and $t\in [0,1].$  
Note that the modes $y_1(t)$ and $y_2(t)$ produce Gaussians in the Fourier domain centered at $k=16$ and $20\Hz$, respectively. The leading term, $me^{-i\pi k}$, centers the wave packets to $t=0.5\s$ where $m=160$ is the total number of samples. For the SRMD algorithm, the hyperparameters to generate the basis are set to $\omega_{\max} = 40$, $N = 20m = 3200$, and $\Delta=0.2$. The hyperparameter for the DBSCAN algorithm is set to $\varepsilon = 1.5$. 

\begin{figure}[htp]
    \centering
    \includegraphics[width=0.4\linewidth]{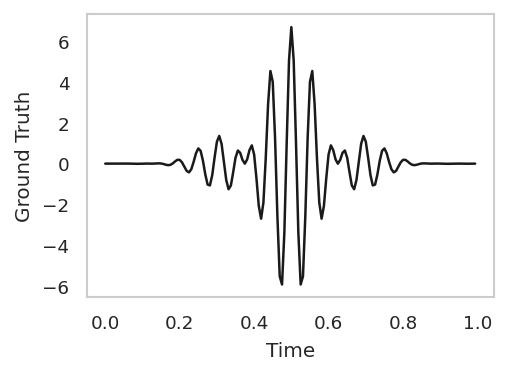}
    \includegraphics[width=0.4\linewidth]{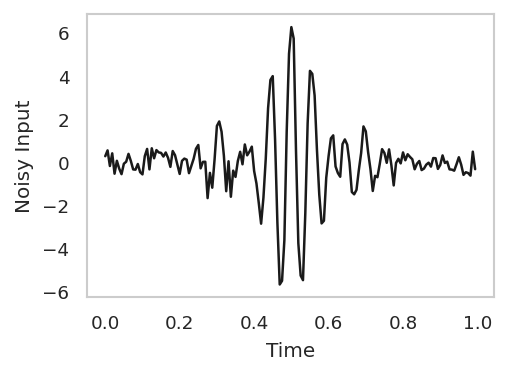}
    \caption{{\bf Example from Section \ref{sec:overlapping}:} Ground truth (left) and noisy input with $r=25\%$ (right).}
    \label{fig:ex_overlapping_25p_noise_inputs}
\end{figure}

The noiseless and noisy time series with $r=25\%$ are shown in Figure \ref{fig:ex_overlapping_25p_noise_inputs}. All reconstruction and decomposition results will be compared against the true signal (or modes) $y_{true}(t), y_1(t),$ and $y_2(t)$. We compare our results with other methods applied to noisy signals with different noise ratios $r = 5\%, 15\%,$ and $25\%$. From the results in Figures \ref{fig:ex_overlapping_5}, we see that only VMD and our method properly reconstruct and decompose the noisy signal. Moreover, when the noise level $r$ is small (5\%), the VMD approach produces comparable results with our method. When $r$ increases, our method is still able to capture the intrinsic modes and denoise the input signal. On the other hand, the VMD is able to identify some aspects of the two intrinsic modes but is polluted by noise, see Figure \ref{fig:ex_overlapping_15} and Figure  \ref{fig:ex_overlapping_25}. 

The clustering of the non-zero coefficients obtained by SRMD applied to the noisy signal with the noise level $r= 5\%, 15\%$, and $25\%$ are shown in Figure \ref{fig:ex_overlapping_clusters}. Note that the clusters surround the two Gaussian peaks ($16 \Hz$ and $20 \Hz$) that define the true input signal. The other features obtained by the SRMD provide slightly corrects to the overall shape to ensure the reconstruction error is below the specified upper bound. This implies one has flexibility in choosing the width and number of random features since additional features can be used to ensure the reconstruction is reasonable.

\begin{figure}[htp]
    \centering
    \includegraphics[width=\figwidth]{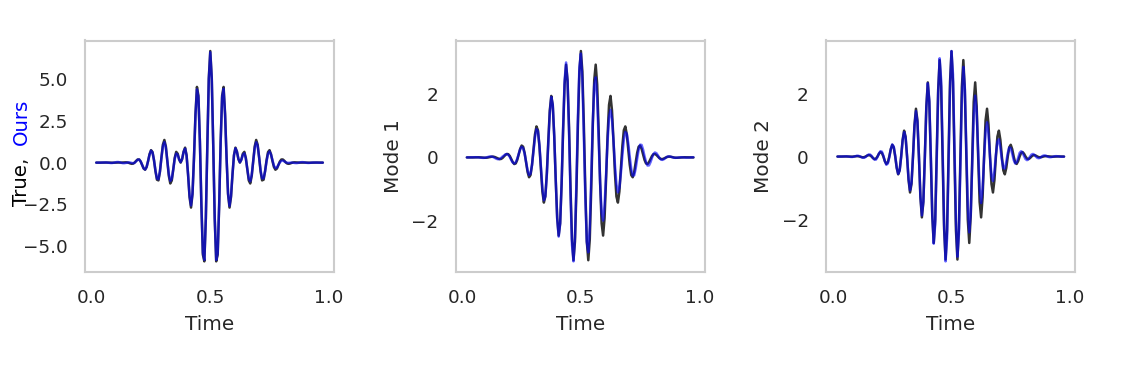}
    \includegraphics[width=\figwidth]{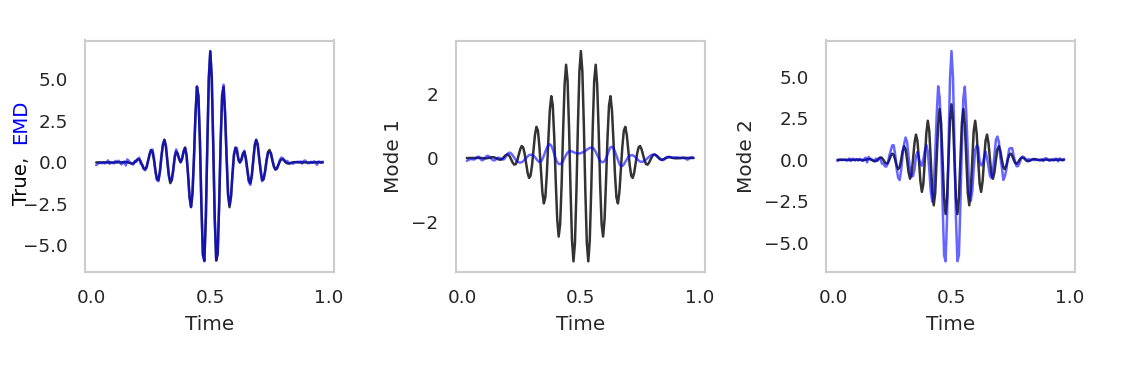}
    \includegraphics[width=\figwidth]{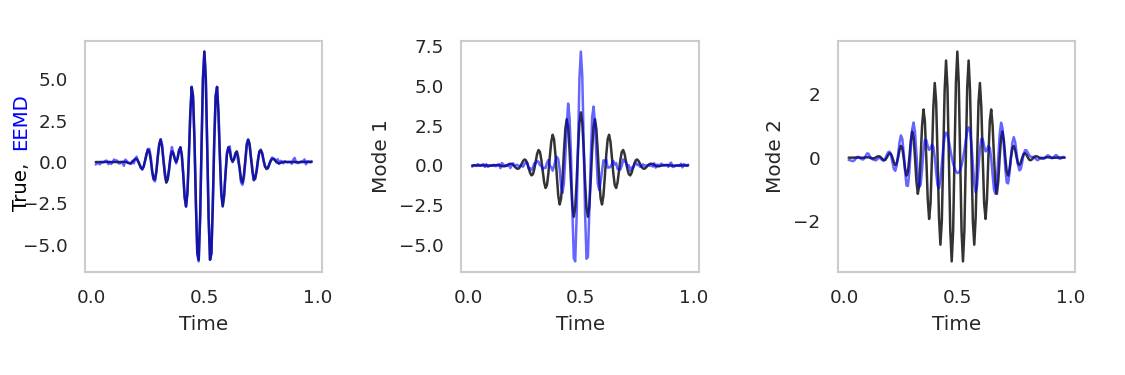}
    \includegraphics[width=\figwidth]{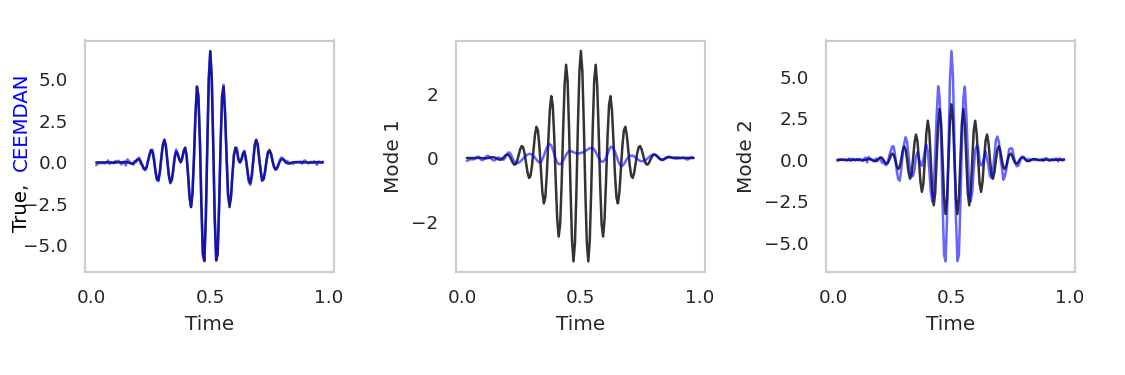}
    \includegraphics[width=\figwidth]{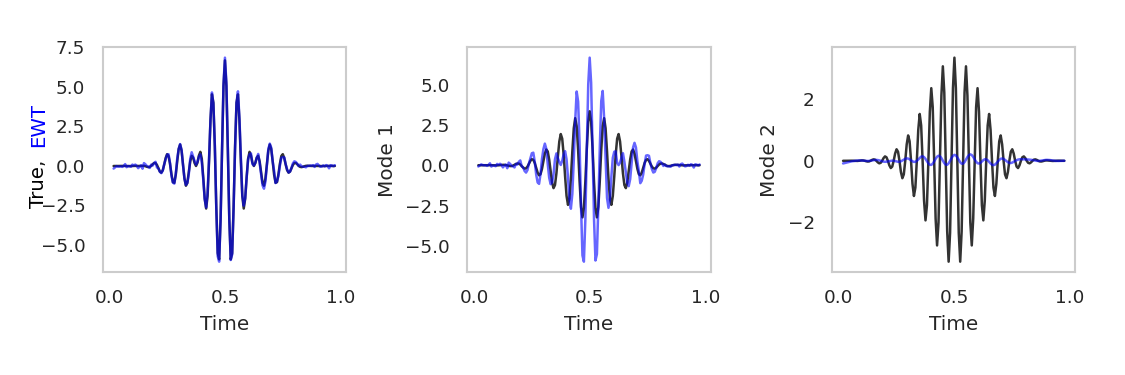}
    \includegraphics[width=\figwidth]{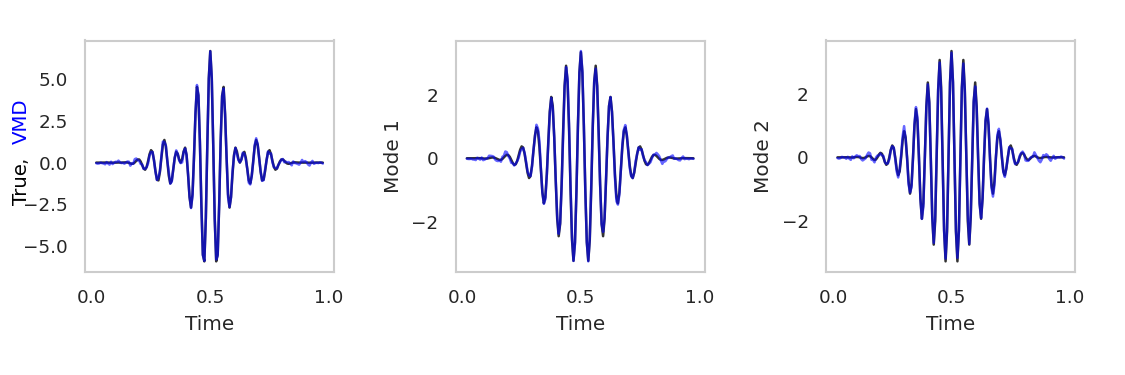}
    \caption{{\bf Example from Section \ref{sec:overlapping}:} Decomposition results with $r= 5\%$ noise using six different methods. Top to bottom rows: SRMD, EMD, EEMD, CEEMDAN, EWT, and VMD. First column: the noiseless ground truth (black) and the learned signal (blue). Middle and last columns: the first and second ground truth IMFs (black) with the learned IMFs (blue).} 
    \label{fig:ex_overlapping_5}
\end{figure}

\begin{figure}[htp]
    \centering
    \includegraphics[width=\figwidth]{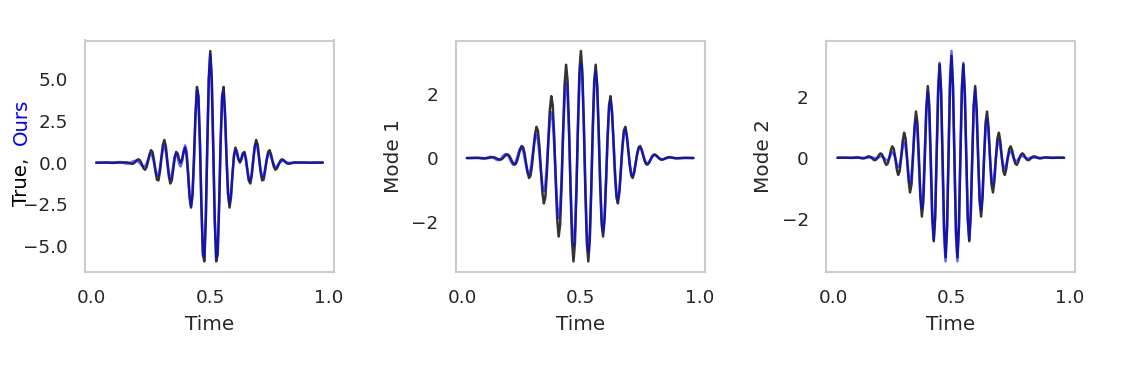}
     \includegraphics[width=\figwidth]{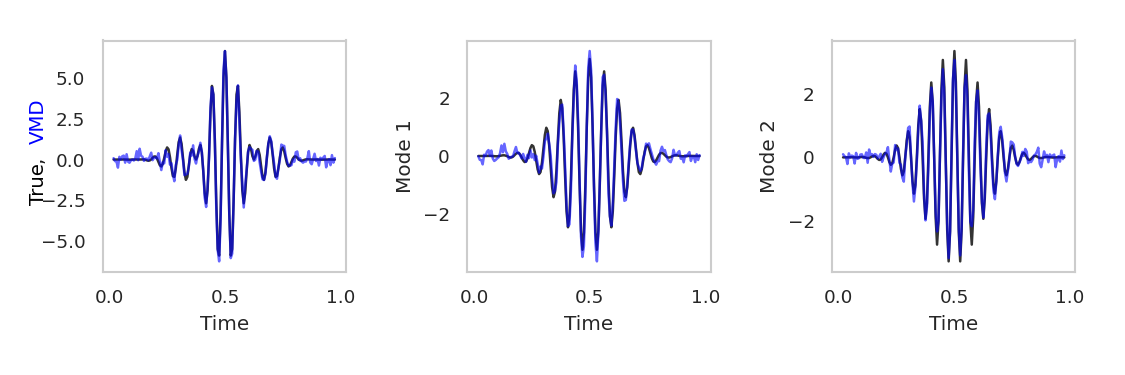}
    \caption{{\bf Example from Section \ref{sec:overlapping}:} Decomposition results with $r= 15\%$ noise using SRMD (first row) and VMD (second row). First column: the noiseless ground truth (black) and the learned signal (blue). Middle and last columns: the first and second ground truth IMFs (black) with the learned IMFs (blue).}
     \label{fig:ex_overlapping_15}
\end{figure}

\begin{figure}[htp]
    \centering
    \includegraphics[width=\figwidth]{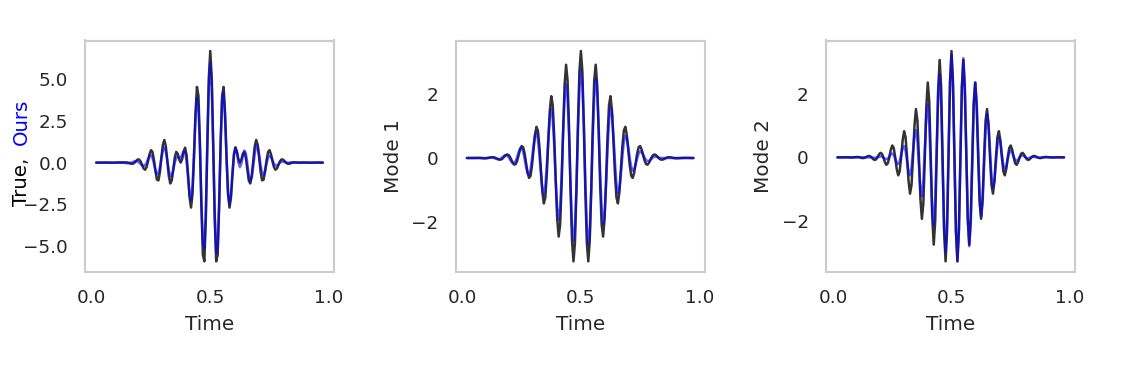}
     \includegraphics[width=\figwidth]{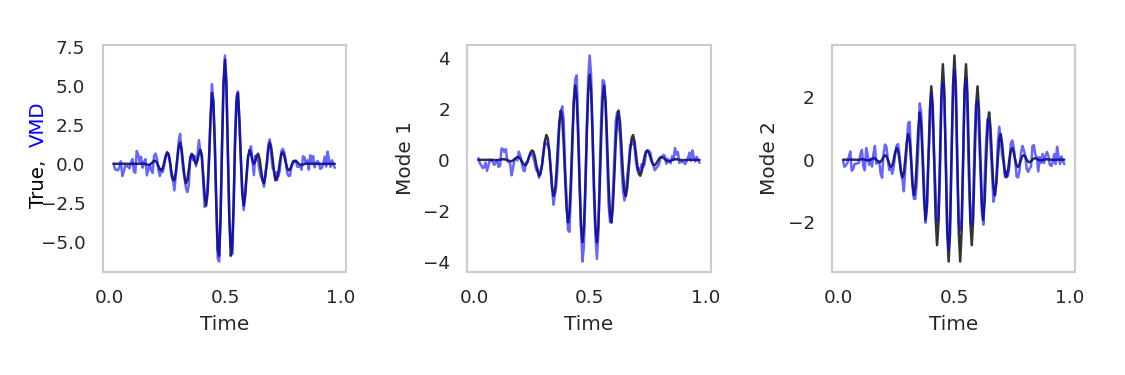}
    \caption{{\bf Example from Section \ref{sec:overlapping}:} Decomposition results with $r= 25\%$ noise using SRMD (first row) and VMD (second row). First column: the noiseless ground truth (black) and the learned signal (blue). Middle and last columns: the first and second ground truth IMFs (black) with the learned IMFs (blue).}
     \label{fig:ex_overlapping_25}
\end{figure}

\begin{figure}[htp]
    \centering
\includegraphics[width = 3.5  in]{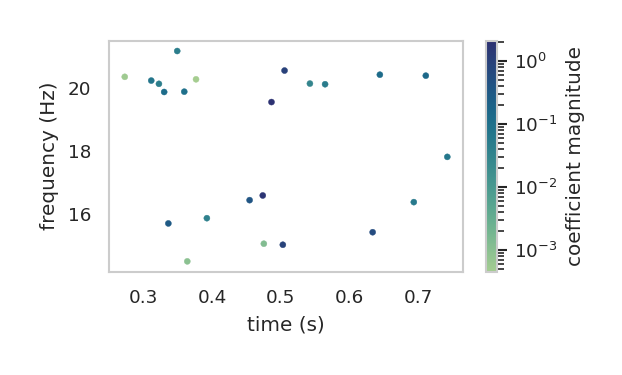}
\includegraphics[width = 2.9 in]{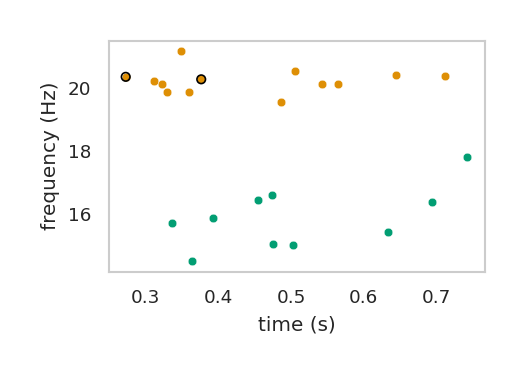}\\
\includegraphics[width = 3.5  in]{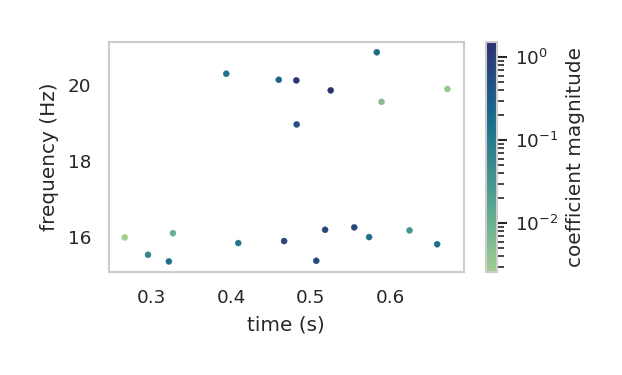}
\includegraphics[width = 2.9 in]{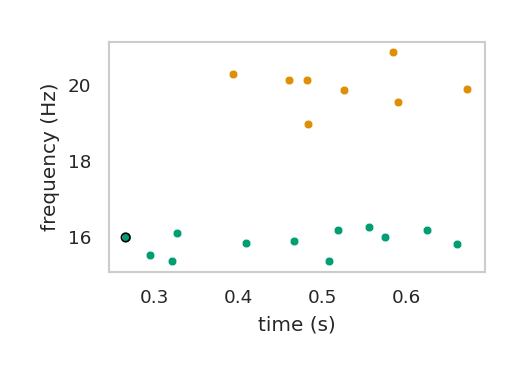}\\
\includegraphics[width = 3.5  in]{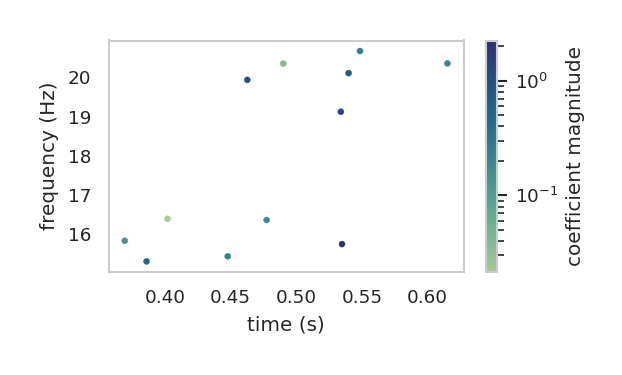}
\includegraphics[width = 2.9 in]{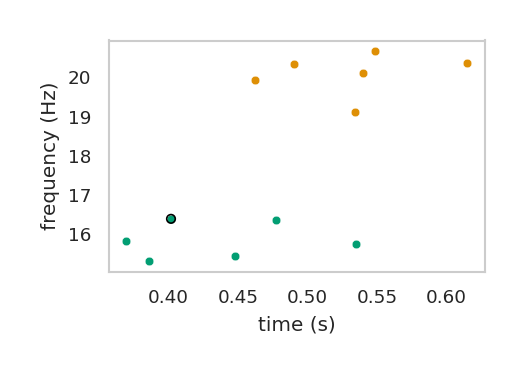}
    \caption{{\bf Example from Section \ref{sec:overlapping}:} First column: Magnitude of non-zero learned coefficients for noisy signals with $r=5\%, 15\%$ and $25\%$. Second column: two learned clusters (green and orange).  }
  \label {fig:ex_overlapping_clusters}
\end{figure}

\subsection{Example on Parameter Tuning and Limitations}\label{sec:limitations}
As we discuss in Section \ref{sec:parameters}, the window size $\Delta$ and the clustering neighborhood scale $\varepsilon$ are sensitive tunable parameters. For example, in the discontinuous time-series example (see Section \ref{sec:discontinuity}), if $\varepsilon = 0.06$ (instead of $0.1$), the non-zero learned coefficients align with the true instantaneous frequencies. However, the learned modes are not reasonable due to wrong clusters (see Figure \ref{fig:learnedWeightsF}). On the other hand, if $\varepsilon = 0.15$, although the non-zero learned coefficients still align with the true instantaneous frequencies, Algorithm \ref{alg:decomposition} can not extract the mode since the clustering part is not able to cluster the set $\widehat{S}$.

\begin{figure}[htp]
    \centering
\includegraphics[width = 2.9  in]{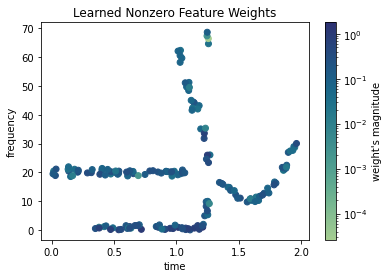}
\includegraphics[width = 2.9 in]{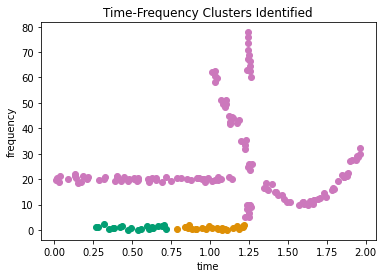}\\
\includegraphics[width = 2.  in]{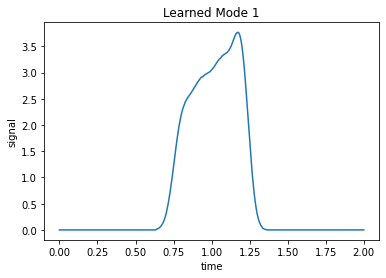}
\includegraphics[width = 2. in]{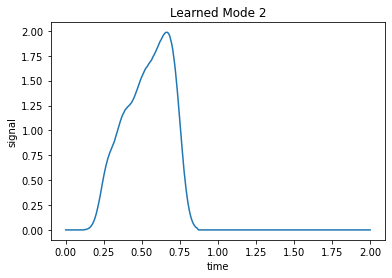}
\includegraphics[width = 2.  in]{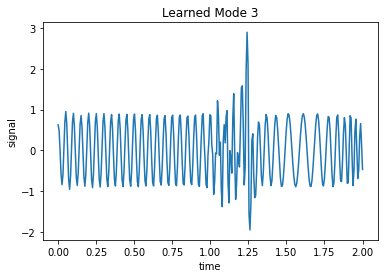}
    \caption{{\bf Example from Section \ref{sec:discontinuity} with DBSCAN hyperparameter $\varepsilon = 0.05$:} First row from left to right: magnitude of non-zero learned coefficients used for DBSCAN and clustering of non-zero coefficients into three modes. Second rom from left to right: three extracted modes.}
  \label {fig:learnedWeightsF}
\end{figure}

Finally, if we choose the window size $\Delta$ too big or too small, the non-zero learned coefficients may not align well with the instantaneous frequencies or make it difficult to cluster (see Figure \ref{fig:DiscontinuityLearnedWeightsF_delta}).
\begin{figure}[htp]
    \centering
    \includegraphics[width = 2.9 in]{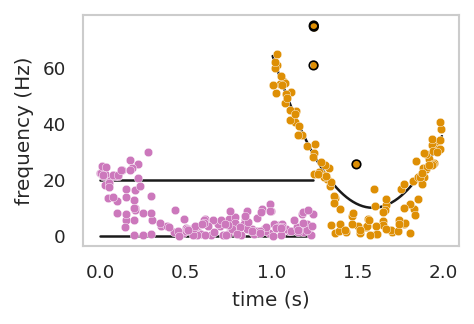}
\includegraphics[width = 2.9 in]{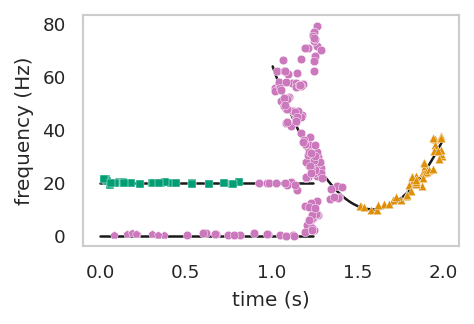}

  \caption{{\bf Example from Section \ref{sec:discontinuity} with various window size $\Delta$:} From left to right: clustering of non-zero coefficients (the true instantaneous frequencies are in black) for $\Delta = 0.02$ (left) and $\Delta = 0.2$.}
 \label{fig:DiscontinuityLearnedWeightsF_delta}
\end{figure}
\end{document}